%% file: main.tex
\documentclass[sigconf,review=False]{acmart}
\AtBeginDocument{%
  }

\setcopyright{acmlicensed}
\copyrightyear{2025}
\acmYear{2026}
\acmDOI{XXXXXXX.XXXXXXX}
\acmConference[AI-SQE]{The 1st International Workshop on AI for Software Quality Evaluation: Judgment, Metrics, Benchmarks, and Beyond}{April 14, 2026}{Rio de Janeiro, Brazil}

\acmISBN{978-1-4503-XXXX-X/2018/06}

\usepackage{longtable}
\usepackage{enumitem}
\usepackage{graphicx}
\usepackage{afterpage}
\usepackage{needspace}
\usepackage{hyperref}
\usepackage{caption}
\usepackage{algorithm}
\usepackage{algpseudocode}
\usepackage{amsmath}
\usepackage{relsize}
\usepackage{tikz}
\usetikzlibrary{shapes.geometric, arrows.meta, shapes.symbols, positioning}
\newcommand{\nth}[1]{#1^{\text{th}}}
\newcommand{\bd}{\bar{\delta}}
\newcommand{\?}{\stackrel{?}{=}}

\newcounter{nodecounter}
\newcommand{\splitatcommas}[1]{\begingroup\lccode`~=`, \lowercase{\endgroup
    \edef~{\mathchar\the\mathcode`, \penalty0 \noexpand\hspace{0pt plus 1em}}%
  }\mathcode`,="8000 #1%
  }

\begin{document}

\title{Evaluating perturbation robustness\\of generative systems that use COBOL code inputs}

\author{Samuel Ackerman}
\authornote{Corresponding author}
\email{samuel.ackerman@ibm.com}
\orcid{0000-0003-2631-0341}

\author{Wesam Ibraheem}

\author{Orna Raz}

\author{Marcel Zalmanovici}

\affiliation{%
  \institution{IBM Research}
  \city{Haifa}
  \country{Israel}
}









\begin{abstract}
Systems incorporating large language models (LLMs) as a component are known to be sensitive (i.e., non-robust) to minor input variations that do not change the meaning of the input; such sensitivity may reduce the system's usefulness.  Here, we present a framework to evaluate robustness of systems using COBOL code as input; our application is translation between COBOL and Java programming languages, but the approach extends to other tasks such as code generation or explanation.  Targeting robustness of systems with COBOL as input is essential yet challenging.  Many business-critical applications are written in COBOL, yet these are typically proprietary legacy applications and their code is unavailable to LLMs for training.  We develop a library of COBOL paragraph and full-program perturbation methods, and create variant-expanded versions of a benchmark dataset of examples for a specific task.  The robustness of the LLM-based system is evaluated by measuring changes in values of individual and aggregate metrics calculated on the system's outputs.  Finally, we present a series of dynamic table and chart visualization dashboards that assist in debugging the system's outputs, and monitoring and  understanding root causes of the system's sensitivity to input variation.  These tools can be further used to improve the system by, for instance, indicating variations that should be handled by pre-processing steps.
\end{abstract}



\maketitle

\input{sections/introduction}
\input{sections/related}
\input{sections/perturbations}

\input{sections/checkers}
\input{sections/experimental_results}

\input{sections/conclusion}

\bibliographystyle{ACM-Reference-Format}
\bibliography{main}

\input{sections/appendix}

\end{document}

%% file: sections/introduction.tex
\section{Introduction 
\label{sec:introduction}}

COBOL (Common Business-Oriented Language) continues to underpin a substantial portion of mission-critical software systems, particularly in domains such as finance, insurance, and government. These systems, often developed and maintained over several decades, embody vast amounts of complex yet highly reliable code. As organizations seek to modernize their software infrastructure, the challenge lies not only in transforming the system implementations but also in evolving the development environments that support ongoing maintenance and enhancement. In this context, generative systems---such as those that translate COBOL to modern languages, generate COBOL code from specifications, or provide automated explanations of COBOL programs---offer promising avenues for modernization. However, the robustness of these systems remains a critical concern, as inaccuracies or inconsistencies can compromise system integrity and developer trust. This work focuses on evaluating the robustness of generative systems that work with COBOL code, aiming to assess their reliability, identify potential failure modes, and inform the development of more resilient tools for legacy system modernization.

Evaluation of such systems can be automated or made easier by using a set of \textbf{evaluation metrics}---typically producing numeric or boolean values---that capture a range of aspects of interest of the system outputs.  For instance, a code-generating system may be assessed by whether the output code is parsable (a boolean measure), the number of variables defined (an integer measure), or by a LLM-as-a-Judge (LaaJ) assessment (typically a real-valued measure) of some aspect of the code, such as its quality.  A system generating a natural-language summary of code may be evaluated by LaaJ scores of the coherence of the summary, or by superficial metrics such as the word count.

In many such generative systems, it is desirable that the generated output be \textbf{robust} to (typically minor) variations (perturbations) in the code inputs.  Assume we have a corpus of example code inputs, for instance from a benchmark dataset. Robustness means that, across inputs in the corpus, the system outputs of perturbations of an input are very similar to the outputs of the original, unperturbed inputs.  In the ideal case, the system outputs themselves will be identical before and after perturbations, but this tends to be too rigid an expectation in reality.  More likely, we will want to see that the values of the \textit{evaluation metrics} measured, and not the outputs themselves, do not change after the perturbation, as our criterion for robustness; exact-output robustness as a special case can be measured by creating a trivial binary evaluation metric comparing the outputs themselves.  Manual inspection of the outputs may be useful for spot-checks, but using measurable metrics is better for evaluating the system. By using enough metrics of interest in assessing the similarity of outputs under perturbation, we hope to provide sufficient experimental guarantees on the robustness of our system in the field. 

To be meaningful, the perturbations should be \textbf{meaning} \textbf{preserving}, a determination which typically requires domain knowledge. 
For instance, in COBOL there is only a two-level distinction in beginning-of-line indentation between Area A (7--10 leading spaces) and B (11 or more leading spaces), such that a line's indentation can be changed as long as it obeys the area's indentation constraint.  In Python, however, there are multiple levels, as each nesting of a function or control flow statements (if, while, for, etc.) requires its own indent.  Likewise, in COBOL, but not Python, non-literal token case can be changed.


Such stylistic or interchangeable grammatical code variations may arise naturally due to varying code practices of individual programmers, or differing company standards.  If a generative translation system is observed to give differing translations on the same input code when the amount of such whitespace changes, there is concern that the procedure is `brittle' and not robust to this variation, and likely others.  Such non-robustness means that the system performance in the field by users may differ from the evaluated performance under development, particularly if the outputs are used downstream.  

The generative task we evaluate here is translation of COBOL to Java code.  We note that our approach and the set of perturbation methods implemented are general enough to be used to evaluate robustness of other tasks such as COBOL code generation or explanation, though we do not address these.  However, robustness evaluation of a task could certainly benefit from inclusion of perturbation methods tailored for that task (e.g., more specific perturbation of comments in code-explanation tasks).

The organization of our paper is as follows: Section ~\ref{sec:related} discusses related work.  Section~\ref{sec:perturbations} discusses the types of perturbations we have implemented, explained in detail in the \hyperref[sec:appendix]{Appendix}. Section~\ref{sec:evaluation} discusses how the degree of perturbation robustness is measured.  Section~\ref{sec:experimental_results} describes the construction of a robustness benchmark dataset and evaluation of it. Section~\ref{sec:conclusion} concludes.

%% file: sections/related.tex
\section{Related work
\label{sec:related}}

The assessment of generative robustness of LLM systems has largely followed the approach of creating expanded datasets of perturbations of inputs, and comparing the system outputs on the original and perturbed inputs.  The evaluation metrics and the perturbations applied depend on the domain and the structure of the system.  In natural language tasks, examples of aspects for which robustness has been evaluated include paraphrasing, editing, and other variants in question-answering (\cite{ackerman2024novelmetricmeasuringrobustness}); premise ordering in logical reasoning (\cite{chen2024premise}); example choice and ordering for in-context learning (ICL) (\cite{voronov2024mind}); and variation in wording of instruction templates (\cite{mizrahi2023stateofwhat}, \cite{sclar2023quantifying}, \cite{zhao2024improving})

For LLM systems with code inputs, the approach of variant-expanded datasets is often used. \cite{chen2024nlperturbator} expand datasets such as HumanEval (\cite{chen2021evaluating}) to evaluate robustness of code generation to phrasing and editing perturbations of function docstrings. \cite{wang2022recode} created ReCode, which similarly conducted natural-language perturbations on function docstrings, renaming the function name in the prompt, or varying the format or syntax of code examples given in the prompt. Here, a robustness version of the pass@k metric from HumanEval is used as the evaluation metric.  \cite{liu2025improving} also perform various editing and paraphrasing perturbations of function docstrings, but use the results to identify and fix non-robust layers in the model itself.

Though we are unaware of similar work on generative robustness with COBOL inputs, there are other works using COBOL with LLMs.  \cite{ali2023xcoboldatasetcobolrepositories} assembled a dataset of COBOL software repositories with source code to, among other uses, study project development.  A COBOL version (\cite{coboleval_repo}, \cite{coboleval_blog}) of the HumanEval dataset was also created.  These datasets could be used as source corpora for LLM performance, evaluation, including for robustness specifically. \cite{lei2025enhancingcobolcodeexplanations} use LLM agents to generate explanations for COBOL code, though no robustness assessment seems to be performed.    

%% file: sections/perturbations.tex
\section{Perturbations
\label{sec:perturbations}}

Here, we discuss the perturbations of COBOL code that we implement.  See \hyperref[sec:appendix]{Appendix} Table~\ref{tab:method_examples} for a full description and examples of inputs and outputs.  As noted in Section~\ref{sec:introduction}, the perturbations are chosen because they are meaning-preserving and cover a range of aspects of observed stylistic and grammatical variations, as suggested by domain experts and client feedback, and are not claimed to be `complete'; additional methods can be added if desired.   
We use the term \textbf{perturbation pipeline} here to refer to the procedure of applying perturbations to a given corpus of input examples (discussed in Sections~\ref{sec:evaluation}, ~\ref{ssec:benchmark_construction}) in a specific way.   These perturbed instances are then run through the same generative system, and then analyzed to measure robustness.

Throughout, we use the notation $X$ to represent input COBOL code text, and $\Delta$ to represent a perturbation method function.  Thus, $X'=\Delta(X)$ is the result of applying method $\Delta$ on $X$.  Further notation is discussed in Section~\ref{ssec:metric_notation}.  By definition, $\Delta(X)$ will always be valid COBOL code that is functionally equivalent to $X$.



\subsection{Methods overview
\label{ssec:methods_overview}}

Each method perturbs input code by performing a single type of modification on a single target feature of the code.  For instance, a target feature may be a comment line, or an identifier name token; an instance of the feature is a particular example of it in the input. Examples of perturbation actions on these features may be to remove the comment line or modify its content, or to change the case of the identifier token.  The perturbed output of the method is the input code where instances of the feature are modified by the action. If there are no feature instances in the input, by definition, the method will have no effect; for instance, if the method is to remove comments, but no comments exist, no change can be caused.  

Some perturbation methods $\Delta$ are \textit{deterministic} in that they take a non-random action on all instances of the feature, for instance removing the content in all comments; $\Delta(X)$ is thus the (unique) result of a single run of $\Delta$.   Others are \textit{probabilistic}, containing a random generator which can govern the probability of the action and/or the result of the action; for instance, a method could toss a coin to determine whether to modify the content of a comment line, then use the generator to determine the random content inserted.  Consider a perturbation that changes the case of an identifier to uppercase: the random coin tossing can be relevant if, for instance, the system output when all the identifiers are in uppercase (deterministic) could differ from when  there is a mix of upper- and lower-case identifiers (probabilistic). Probabilistic methods are allowed up to 30 tries\footnote{Assuming the random likelihood on a given instance of affecting change is not nearly 0, in which case $\Delta$ is not very useful by design, 30 tries is more than sufficient to obtain at least one change in at least one instance across the tries.
}---without resetting the generator---to affect a change in the input code, the first such success being the output code $\Delta(X)$.

Below we list the perturbation categories; the number in parentheses is the number of sub-methods.  The first categories are \textit{syntactic}, in that they use domain knowledge of the language grammar:

\begin{itemize}[leftmargin=10px]  
    \item \textbf{end-block} (6): the token \texttt{END-{XXX}} is used to indicate the end of particular blocks, such as IF and EVALUATE statements.  In certain cases, either 1) \texttt{END-{XXX}} when not followed by a period may be removed and replaced with a period, or 2) \texttt{END-{XXX}} when followed by a period may be removed, leaving the period.
    \item \textbf{within-line whitespace} (4): variations in between-token whitespace and leading and trailing whitespace within an individual line, while maintaining the Area A/B assignment of lines and allowed line length.  
    \item \textbf{line breaks} (5): combining successive lines, or breaking lines into two; the latter case includes continuation-line perturbations discussed in Section~\ref{ssec:free_fixed_cobol}.
    \item \textbf{distraction} (1): insertion of a code token that does not affect the code.
    \item \textbf{empty lines} (3): insertion of empty lines between \textit{existing} lines 
    (not line breaks), or removal of empty lines.
    \item \textbf{comments} (3): removal or modification of comment line contents.
    \item \textbf{case} (4): change of case of non-literal tokens, in particular identifier names.
\end{itemize}

The final two categories conduct global random \textit{remaming} of identifiers (other than special registers or fixed identifiers associated with copybooks) in the data division.  They can only be applied on whole programs, while the syntactic ones can receive sub-units such as sections or paragraphs:
    
    \begin{itemize}[leftmargin=10px]  
        \item \textbf{identifier-renaming-obfuscation} (4): changing the identifier name in a way that breaks any connection between the name and the identifier (e.g., a gibberish name).
        \item \textbf{identifier-renaming-preserve} (3): a slight modification that preserves the essence of the name.
    \end{itemize}

In theory, we might want a generative system to be robust to, say, identifier name changes, in that it wouldn't use the information `clues' inherent in the names in being able to translate or summarize the program.  However, human coders---whose code would typically be used in training the system model---usually give descriptive names to functions, parameters, identifiers, and other objects, to facilitate human comprehension.  Thus, the model is likely to learn a relationship between standardized names and the output, based on the training examples.  The obfuscation method types test whether the system is robust to complete loss of the information in the identifier name, while the `preserve' types test robustness to slight changes that preserve most of the name information.  We would expect worse robustness for obfuscation.

\subsection{Free- and fixed-form COBOL
\label{ssec:free_fixed_cobol}}

We note also that COBOL code can be either in \textit{free-form} or \textit{fixed-form}.  The primary differences are
\begin{itemize}[leftmargin=10px]  
    \item Line length: in fixed-form, lines must be at most 72 characters long (including leading and trailing whitespace).  Continuation line indicators\footnote{The rules for where continuation line splits are allowed are complex and listed in \cite{cobol-lang-ref} the \href{https://www.ibm.com/docs/en/cobol-zos/6.5.0?topic=b-continuation-lines}{reference}.} can be used, where legal, to split code---typically a literal---between two or more lines, especially if the length limit would be violated otherwise.   In free-form, the line length is not restricted, and continuation lines are not used. 
    \item Comment format: Comment lines begin with a single asterisk in fixed-form, and with \texttt{*>} in free-form.
\end{itemize}

COBOL code can be converted between fixed- and free-form.  Aside from the comment format, converting fixed- to free-form (`fixed2free') code involves joining a block of continuation lines into a single line while, if the split is not within a literal,  discarding leading whitespace after the continuation line.   Also, fixed2free modifies commas or semicolons followed by newlines (e.g., a comma-separated list of identifiers that was formatted by inserting newlines between items) by appending a single space before the newline.  These modifications thus involve a loss of information, in particular with continuation lines since, in free-form, the original configuration of splits of a given line---or even \textit{if} it was originally split, which is optional if the freeform line is less than 72 characters long---is lost.  Thus fixed2free is a many-to-one mapping since two fixed-form programs that differ only in the locations of continuation splits can be converted to the same free-form program.  

Conversely, mapping a free-form program to fixed-form is potentially one-to-many because the placement of continuation lines can be varied.  Here, we use `standard fixed form' (SFX) to refer to the result of free2fixed conversion of a free-form COBOL line where only a line longer than 72 characters is split; for that line, each continuation line segment is created by splitting at the highest legal location---so that each split is as close to 72 characters as possible, given the previous ones---and with four leading spaces after the continuation line hyphen. Converting a free-form program to SFX form involves changing comment line format, and applying SFX form on all other lines in the program.  Our contribution includes an implementation of an SFX free2fixed conversion utility.

Two of the defined perturbation methods under the `line breaks' category return non-standard (i.e., not SFX) fixed-form output: 1) forcing, if possible, a continuation split at a random location in each literal, rather than at the maximal length, or 2) splitting at random locations in lines, with a coin toss.  Both perturbations are done in some cases even when the free-form line is not long enough to require the split, which is what makes them perturbations.

\subsection{Perturbation pipeline 
\label{ssec:perturbation_pieline}}

In our perturbation pipeline, the COBOL input and desired output formats can each be either fixed- or free-form (user-indicated); the default assumption is that both the input and output are fixed-form.  The
pipeline can be run for any combination of
perturbation method $\Delta$, COBOL input $X$ (which determines whether the input is free- or fixed-form), and desired output format (free or fixed).  The exception is that $\Delta$ cannot be one of the non-SFX continuation line perturbations above when free-form output---which cannot have continuation lines---is desired.

Figure~\ref{fig:flowchart} illustrates our pipeline procedure given these three inputs.  Green banner nodes represent decisions, orange cylinder nodes are processes, and white rectangular terminal nodes are the (potentially) perturbed outputs.  The three possible returned results when both input and output are fixed-form (the default) are shown as white nodes 9, 13, and 18 with thick borders in Figure~\ref{fig:flowchart}.

Certain perturbations $\Delta$, such as between-token splits or end-block changes, require identification of grammatical elements; this can only be done by parsing the COBOL code in \textit{free-form}, necessitating fixed2free conversion if the original code is in fixed-form.  In this case, node 3 is entered.  The perturbation $\Delta$ causes a change if the free-form version of $X$ changes after applying $\Delta$ (node 6).

LLM-based generative systems---in contrast to more explainable ML-based systems---can often behave in unexplainable ways, in part due to learned artefacts in the training corpus.  And since here we are investigating generative robustness to meaning-preserving perturbations---many of which are `superficial', such as varying whitespace, case, or newlines---it is crucial to avoid introducing superficial `contaminations' as a by-product of the perturbation pipeline, as this would mean we are not measuring only the effect of the perturbation method itself.

We noted that free-form conversion can cause a loss of information in $X$ (Section~\ref{ssec:free_fixed_cobol}), such as in the specific placement of continuation line breaks. Thus, whenever first parsing in free-form is not required, perturbation methods are implemented to be applied on the original $X$, with slight adaptations depending on whether it is in fixed- or free-form.  For instance, a perturbation of inserting empty lines, which doesn't need parsing, can work on fixed-form if we ensure the inserted empty lines don't come in the middle of a block of continued lines; this modification is not needed if $X$ was already in free-form, where empty lines can be inserted anywhere.  In this case, the perturbed output $X'$ can be compared directly to $X$ in its original form (node 11), without first converting to free-form.  

For elaboration, we give counter-examples where contaminations could be introduced that are not removed.  Assume throughout that $X$ is in fixed-form and we want fixed-form output; say that the the method $\Delta$ is to convert tokens to uppercase, which first requires free-form conversion of $X$ to parse it. First, say we have input code $X$ that is in non-SFX form, that is, there are continuation breaks in different locations than we would get if the input underwent fixed2free conversion then back to SFX via our free2fixed utility.  We then attempt the uppercase token perturbation, which requires free-form conversion.  If the perturbation doesn't have an effect (e.g., all tokens are already in uppercase), then the original $X$ (in non-SFX form) is returned.  If the perturbation does have an effect, free2fixed conversion is applied on the changed free-form result.  Because the final perturbed $X'$ is now in SFX form, in addition to the token case changes we will see changes in the non-SFX continuation lines in $X$ that now have different breaks, if any, in $X'$ which is in SFX form.  That is, the result contains changes that are not solely the result of the perturbation.  In this case, maintaining the best fidelity to the original $X$ (i.e., keeping the non-SFX lines in their original form in $X'$) would practically be too cumbersome to attempt to do.  

As mentioned, fixed2free conversion also inserts a single space after commas and semicolons if followed immediately by a newline.  Say fixed-form $X$ now is in SFX form but has cases where newlines immediately follow commas.  If the token case perturbation is applied, the result $X'$, if token case changes can be made, will also have spaces after these commas when converted to fixed-form, which are not a direct result of the case perturbation, but which could be difficult to undo.  If not, the original $X$ is returned.


\tikzstyle{io} = [rectangle, rounded corners, text width=2cm, minimum height=0.5cm,text centered, draw=black, fill=red!30]
\tikzstyle{process} = [cylinder, aspect=0.25, text width=2cm, minimum height=0.5cm, text centered, draw=black, fill=orange!30]
\tikzstyle{return} = [rectangle, text width=2cm, minimum height=0.5cm, text centered, draw=black]
\tikzstyle{decision} = [tape, tape bend top=none, text width=2cm, minimum height=0.5cm, text centered, draw=black, fill=green!30]
\tikzset{>={Stealth[scale=2]}}
\tikzset{node distance = 3cm and 3cm}

\begin{figure}
\resizebox{\columnwidth}{!}{%
\begin{tikzpicture}
\stepcounter{nodecounter}
\node (start) [io, label=above right:\arabic{nodecounter}] {COBOL code $X$, perturbation method $\Delta(\cdot)$};
\stepcounter{nodecounter}
\node (req_parse) [decision, below of=start, node distance=2.5cm, label=above right:\arabic{nodecounter}] {$\Delta$ requires free-form parse?};
\stepcounter{nodecounter}
\node (input_is_freeform) [decision, below left of=req_parse, label=above left:\arabic{nodecounter}] {is $X$ free-form};
\stepcounter{nodecounter}
\node (apply_fixed2free) [process, left of=input_is_freeform, node distance=4cm, label=above right:\arabic{nodecounter}] {free($X$) = fixed2free($X$)};
\stepcounter{nodecounter}
\node (parse_apply) [process, below of=apply_fixed2free, label=above left:\arabic{nodecounter}] {Apply $X'=\Delta(\textrm{free}(X))$};
\stepcounter{nodecounter}
\node (parse_is_unchanged) [decision, below of=parse_apply, line width=3pt, label=above right:\arabic{nodecounter}] {$X'\?\textrm{free}(X)$};
\stepcounter{nodecounter}
\node (parse_changed_which_want) [decision, below left of=parse_is_unchanged, label=above left:\arabic{nodecounter}] {want free-form output?};
\stepcounter{nodecounter}
\node (parse_changed_want_free) [return, below left of=parse_changed_which_want, label=above left:\arabic{nodecounter}] {return $X'$};
\stepcounter{nodecounter}
\node (parse_changed_want_fixed) [return, below right of=parse_changed_which_want, line width=3pt, inner sep=7pt, label=above right:\arabic{nodecounter}] {return free2fixed($X'$)};

\stepcounter{nodecounter}
\node (no_parse_apply) [process, below right of=req_parse, label=above right:\arabic{nodecounter}] {Apply $X'=\Delta(X)$};
\stepcounter{nodecounter}
\node (no_parse_is_unchanged) [decision, below of=no_parse_apply, line width=3pt, label=above right:\arabic{nodecounter}] {$X'\?X$};
\stepcounter{nodecounter}
\node (no_parse_changed_is_input_output_same) [decision, below right of=no_parse_is_unchanged, label=above right:\arabic{nodecounter}] {$X$ input form and desired output form are same?};
\stepcounter{nodecounter}
\node (no_parse_changed_same_form_output) [return, below left of=no_parse_changed_is_input_output_same, line width=3pt, label=above left:\arabic{nodecounter}] {return $X'$};
\stepcounter{nodecounter}
\node (no_parse_changed_diff_form_output_what_is_X) [decision, below right of=no_parse_changed_is_input_output_same, label=above right:\arabic{nodecounter}] {is $X$ free-form?};
\stepcounter{nodecounter}
\node (no_parse_changed_diff_form_output_want_free) [return, below left of=no_parse_changed_diff_form_output_what_is_X, inner sep=7pt, label=above left:\arabic{nodecounter}] {return free2fixed($X'$)};
\stepcounter{nodecounter}
\node (no_parse_changed_diff_form_output_want_fixed) [return, below right of=no_parse_changed_diff_form_output_what_is_X, inner sep=7pt, label=above right:\arabic{nodecounter}] {return fixed2free($X'$)};

\stepcounter{nodecounter}
\node (unchanged_is_input_output_same) [decision, left of=no_parse_is_unchanged, node distance=3cm, label=above right:\arabic{nodecounter}] {$X$ input form and desired output form are same?};
\stepcounter{nodecounter}
\node (unchanged_same_form_output) [return, below left of=unchanged_is_input_output_same, line width=3pt, node distance=4cm and 2cm, label=above left:\arabic{nodecounter}] {return $X$};
\stepcounter{nodecounter}
\node (unchanged_diff_form_output_what_is_X) [decision, below of=unchanged_is_input_output_same, node distance=3.5cm, label=above right:\arabic{nodecounter}] {is $X$ free-form};
\stepcounter{nodecounter}
\node (unchanged_diff_form_output_x_is_free) [return, below left of=unchanged_diff_form_output_what_is_X, inner sep=7pt, label=above left:\arabic{nodecounter}] {return free2fixed($X$)};
\stepcounter{nodecounter}
\node (unchanged_diff_form_output_x_is_fixed) [return, below right of=unchanged_diff_form_output_what_is_X, inner sep=7pt, label=above right:\arabic{nodecounter}] {return fixed2free($X$)};

\draw [->] (start) -- (req_parse);
\draw[->] (req_parse) --node {yes} (input_is_freeform);
\draw[->] (req_parse) --node {no} (no_parse_apply);

\draw[->] (input_is_freeform) --node[midway, above] {no} (apply_fixed2free);
\draw[->] (input_is_freeform) -- (parse_apply) node[midway, above, align=center] {yes\\(free($X$)=$X$)} ;
\draw[->] (apply_fixed2free) -- (parse_apply);
\draw[->] (parse_apply) -- (parse_is_unchanged);
\draw[->] (parse_is_unchanged) -- node {no} (parse_changed_which_want);
\draw[->] (parse_changed_which_want) -- node {yes} (parse_changed_want_free);
\draw[->] (parse_changed_which_want) -- node {no} (parse_changed_want_fixed);

\draw[->] (parse_is_unchanged) --node {yes} (unchanged_is_input_output_same);
\draw[->] (unchanged_is_input_output_same) -- node {yes} (unchanged_same_form_output);
\draw[->] (unchanged_diff_form_output_what_is_X) --node  {yes} (unchanged_diff_form_output_x_is_free);
\draw[->] (unchanged_diff_form_output_what_is_X) --node  {no} (unchanged_diff_form_output_x_is_fixed);

\draw[->] (no_parse_apply) -- (no_parse_is_unchanged);
\draw[->] (no_parse_is_unchanged) -- node[midway, above] {yes} (unchanged_is_input_output_same);
\draw[->] (unchanged_is_input_output_same) -- node {no} (unchanged_diff_form_output_what_is_X);
\draw[->] (no_parse_is_unchanged) -- node {no}(no_parse_changed_is_input_output_same);
\draw[->](no_parse_changed_is_input_output_same) -- node {yes} (no_parse_changed_same_form_output);
\draw[->](no_parse_changed_is_input_output_same) -- node {no} (no_parse_changed_diff_form_output_what_is_X);
\draw[->](no_parse_changed_diff_form_output_what_is_X) -- node {yes} (no_parse_changed_diff_form_output_want_free);
\draw[->](no_parse_changed_diff_form_output_what_is_X) -- node {no} (no_parse_changed_diff_form_output_want_fixed);
\end{tikzpicture}%
} 
\caption{\label{fig:flowchart} Flowchart of perturbation pipeline for an input code example $X$ and desired perturbation method $\Delta$.}
\end{figure}
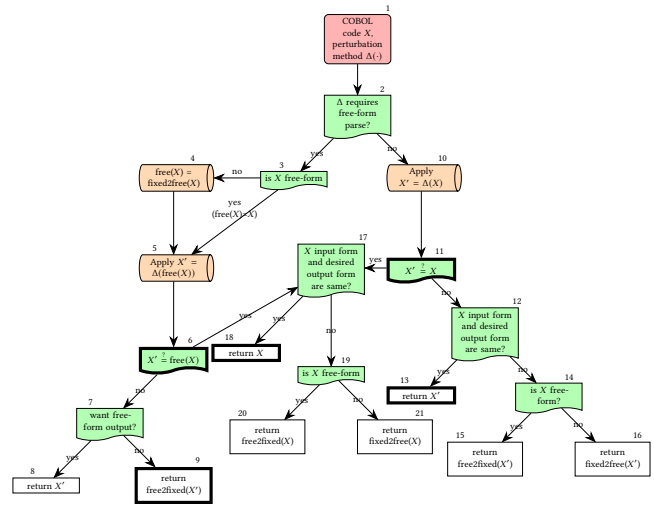

%% file: sections/checkers.tex
\section{Evaluation metrics
\label{sec:evaluation}
}

\subsection{Notation
\label{ssec:metric_notation}
}

To evaluate the robustness of the generative process, a given input is perturbed according to multiple methods (Section~\ref{sec:perturbations}), and the output of the perturbed input is compared to that of the original input.  Sometimes, superficial aspects, such as character counts, can be used to compare the outputs.  In our application, however, we find it more informative to measure changes in non-superficial aspects of interest, such as its parsability. 

As in Section~\ref{sec:perturbations}, $X$ denotes input COBOL code text and $\Delta(\cdot)$ is a perturbation method function applied to it.  In addition, let $\mathcal{D}(X)$ be the set of valid perturbation method functions that can be applied to $X$; for instance, a fixed-form continuation lines perturbation cannot be applied to $X$ if free-form output is desired, so $\mathcal{D}(X)$ excludes such methods.  Thus, we must have $\Delta\in\mathcal{D}(X)$ for $\Delta(X)$ to be valid.

Since robustness evaluation requires variation to compare to a baseline, we will always assume that $\Delta(X) \ne X$ in the resulting dataset expansion (Section~\ref{ssec:benchmark_construction}).  That is, when a perturbation generation procedure is used, we omit instances of $\Delta$s where $\Delta(X) = X$, i.e., no actual perturbation was done; for instance, if method $\Delta$ removes empty lines but no such lines exist, that instance $\Delta(X)$ for $X$ is ignored, and a different method $\Delta'$ is used.

Let $f(\cdot)$ denote the result of the generative system.  For instance $f(X)$ may be the generated translation of input code $X$ to a different language, or a natural-language summary.

Let $\mu(\cdot)$ be a function (e.g., `checkers' as described in \cite{froimovich2025quality}) that measures a certain aspect of the generative output.  In particular, the measurement may be boolean- (e.g., whether the translated code is parsable or not) or numeric-valued (e.g., a count of the number of identifiers appearing in the translated code, or a LaaJ score of it).

\subsection{Measuring robustness
\label{ssec:measuring_robustness}}

Our robustness evaluation framework involves observing how often perturbations change the measured aspect of a generated output.  Thus, we will compare $\mu(f(X))$ vs $\mu(f(\Delta(X))$, across inputs $X$, for multiple perturbations $\Delta$, and multiple metrics $\mu$.

Table ~\ref{tab:perturbed_table} shows an illustration of the results of perturbations.  Imagine we have a corpus $\mathbf{X}=(X_1,X_2,\dots,X_N)$ of $N$ code inputs $X_i$.  Each horizontal rectangular block (group index $i\in1,\dots,N$) corresponds to an input $X_i$ (the first row in the block) and $n(i) \geq 1$ perturbations of it.  Here $n(1)=n(2)=3$ but the number need not be fixed across $i$.  Within a block $i$, $\Delta_{i,j}$ is the $\nth{j}$ (of $n(i)$) perturbation method applied to input $X_i$, with output perturbed code $\Delta_{i,j}(X_i)$.  Typically, for each $i$, $(\Delta_{i,1},\dots,\Delta_{i,n(i)})$ are unique, that is, $n(i)$ \textit{different} perturbation methods were applied to $X_i$.

Not shown in the table is $f$, the system output of the `input' column values (e.g., the code translation).  In this example, $M=3$ evaluation criteria metrics (`checkers') are used on the output $f$.  $\mu_1$ and $\mu_2$ have numeric output, while $\mu_3$ is boolean.  For each metric $\mu_m$, $\delta_m$ is a binary indicator of whether the value changed relative to the unperturbed.  Formally, $\delta_{i,j,m}=I\left(\mu_m(f(\Delta_{i,j}(X_i))) \:\ne\: \mu_c(f(X_i))\right),\:j=1,\dots,n(i)$.  The final column is $\delta_{i,j}=\delta_{i,j,1} \lor \dots \lor \delta_{i,j,M}$, a binary indicator of whether, for the $\nth{j}$ perturbation of input $X_i$, \textit{any} of the metric values changed.

Note that mathematically, $\mu$ and $f$ are functions.  Thus, $(\delta_{i,j,m} = 1) \Rightarrow f(\Delta_{i,j}(X_i)) \ne f(X_i)$, but it is not true that $(\delta_{i,j,m} =  0) \Rightarrow f(\Delta_{i,j}(X_i)) = f(X_i)$.  That is, a change in the metric value $\mu$ means there was a change in the system output $f$, but the output $f$ can (frequently) change without a change in $\mu$.

\begin{table}[ht]
\begin{center}
\hspace*{-4mm}\begin{tabular}{l| l| l l  l| l l l |l} 
\hline
Grp. & Input & $\mu_1(f)$ & $\mu_2(f)$ & $\mu_M(f)$ & $\delta_1$ & $\delta_2$ & $\delta_M$ & $\delta$ (any)\\
\hline
1 & $X_1$ & 1 & 5 & True & - & - & - & -\\
1`& $\Delta_{1,1}(X_1)$ & 1 & 5 & True & 0 & 0 & 0 & 0\\
1 & $\Delta_{1,2}(X_1)$ & 2 & 5 & False & 1 & 0 & 1 & 1\\
1 & $\Delta_{1,n(1)}(X_1)$ & 1 & 4 & True & 1 & 0 & 0 & 1\\
\hline
2 & $X_2$ & 3 & 7 & False  & - & - & -& -\\
2`& $\Delta_{2,1}(X_2)$ & 3 & 6 & False & 0 & 1 & 0 & 1\\
2 & $\Delta_{2,2}(X_2)$ & 3 & 5 & True & 0 & 1 & 1 & 1\\
2 & $\Delta_{2,n(2)}(X_2)$ & 3 & 7 & False & 0 & 0 & 0 & 0\\
\hline
$\vdots$ & $\vdots$ & $\vdots$  & $\vdots$ & $\vdots$ & $\vdots$  & $\vdots$ & $\vdots$  & $\vdots$\\
\hline
\end{tabular}
\end{center}
\caption{\label{tab:perturbed_table} Representative table of perturbed instances.}
\end{table}

\subsection{Robustness proxy measures
\label{ssec:robustness_proxy_measures}
}

For an evaluation metric (checker) $\mu_m$, we can define its rate of change on the $\nth{i}$ input $X_i$ and its perturbations as $\bd_{i,\cdot,m}=\frac{\sum_{j=1}^{n(i)}\delta_{i,j,m}}{n(i)}$.  Its dataset-average rate of change is $\bd_m=\frac{\sum_{i=1}^N\bd_{i,\cdot,m}}{N}$.  This serves as an (anti)-robustness proxy measure for the metric because it indicates how often on average it changes when the input is perturbed, over the distribution of inputs $X_i$ in the dataset, which are assumed to be representative.  The average `any metric changed' rate can be likewise be calculated as $\bd=\frac{1}{N}\mathlarger{\sum}_{i=1}^N \left(\frac{\sum_{j=1}^{n(i)}\delta_{i,j}}{n(i)}\right)$, using the variant-level any-metric change indicator $\delta_{i,j}$ defined previously.  We propose using $\bd$ as a proxy measure for robustness for the input corpus $\textbf{X}$ and the set of perturbations considered, that is conditioned on the set $M$ of metrics.  Naturally, as the number $M$ of metrics used increases, the likelihood of any of them changing increases.  The $M$ metrics should be chosen experimentally to reflect the important aspects of the system output $f$, so that a change in their values after perturbation of inputs $X$ could indicate significant deviation in the output value $f$.  Using a metric $\mu_m$ whose change indicators $\delta_{\cdot,\cdot,m}$ tend to frequently be 1 suggests either that the aspect measured is too superficial (e.g., the character length of the output, nearly as sensitive as the value of the $f$ itself), or if not, that the system itself simply has very poor robustness.   

We can also calculate the \textit{perturbation-conditional robustness}.  Let $D\subseteq\mathcal{D}(X)$ be a non-empty unique set of perturbation methods.  Of particular interest will be if $|D|=1$ (a single method $\Delta$, e.g., concatenation of two consecutive lines into a longer line) or if $D$ represents a category of similar methods (Section~\ref{ssec:methods_overview}). 

The average $D$ method(s)-conditional robustness is calculated as follows.  Recall that the dataset Table~\ref{tab:perturbed_table} excludes instances of methods that had no effect on the input $X_i$. Let $v(i)=\sum_{j=1}^{n(i)}I(\Delta_{i,j} \in D)$ be the number out of $n(i)$ variants of $X_i$ that were perturbed by one of the methods in $D$.  When $|D|=1$, typically $v(i)\in \{0,1\}$ because the number of perturbations performed $n(i)$ is low relative to the number of available methods $|\mathcal{D}(X_i)|$.  Let $\mathcal{I}_D=\{i\colon v(i)\geq 1,\: i=1,\dots,N\}$ be the indices of inputs $X_i$ where one of the methods in $D$ was applied at least once.  Finally, let 
$\bd_D=\frac{1}{|\mathcal{I}_D|}\mathlarger{\sum}_{i\in \mathcal{I}_D}\left(\frac{\sum_{j=1}^{n(i)}I(\Delta_{i,j}\in D)\delta_{i,j}}{v(i)}\right)$.
That is, for each input $X_i$ on which a method in $D$ was applied at least once ($i\in \mathcal{I}_D$), we calculate the average probability of the method changing any metric ($\delta_{i,j}=1$) for only those $n(i)$ instances.  This rate is then averaged across all relevant inputs $X_i,\:i\in\mathcal{I}_D$.
The corresponding rate at which a method in $D$ changes a particular metric $m$ can be calculated as $\bd_{D,m}=\frac{1}{|\mathcal{I}_D|}\mathlarger{\sum}_{i\in \mathcal{I}_D}\left(\frac{\sum_{j=1}^{n(i)}I(\Delta_{i,j}\in D)\delta_{i,j,m}}{v(i)}\right)$.

When a method or set of methods $D$ has higher $\bd_D \in [0,1]$, the system is less robust to these methods, conditional on them being applied; if the method is infrequently occurring in the field in practice, however, this may not be a concern.

%% file: sections/experimental_results.tex
\section{Experimental results
\label{sec:experimental_results}
}

\subsection{Benchmark Construction
\label{ssec:benchmark_construction}
}

We demonstrate our perturbation pipeline on a COBOL-to-Java (C2J) code translation system \cite{froimovich2025quality}, on the \textit{genapp21} dataset described there.\footnote{The perturbation pipeline was also performed on \textit{cblstmts}, a larger internally-curated dataset of 21 programs containing 244 paragraphs, that attempts to cover all clause and sub-clause variations of the COBOL language.}  The C2J system conducts translation at the level of \textit{paragraphs} extracted from a source corpus of COBOL programs.  The \textit{genapp21} dataset $\mathbf{X}$ consists of $N=22$ paragraphs $(X_1,\dots,X_{22})$ extracted from 9 COBOL programs.  The analysis ignores any additional paragraphs in the program corpus that are not extracted for evaluation.

Our pipeline creates perturbed \textit{full programs}; in the C2J task, only the relevant paragraphs and their variants are extracted, but for other tasks, such as code explanation, the full program may be used.  We create a robustness benchmark dataset expanded from the extracted paragraphs, as illustrated in Table~\ref{tab:perturbed_table}.  For each program, $R_r\geq 1$ renaming and $R_s\geq 1$ syntactic variants are created.

The $R_r$ renaming variants are chosen randomly, if possible without replacement; each paragraph in the same program variant undergoes the same renaming perturbation because the program must be internally consistent.  For syntactic perturbations, each paragraph in each of the $R_s$ variants of a program is perturbed independently until a successful change, yielding a mix of methods in a program, while attempting to avoid repeating perturbation methods across the $R_s$ variants of a paragraph across programs.  Thus for each paragraph $X$, the syntactic perturbation methods thus would appear roughly in proportion---according to  repeated sampling \textit{without} replacement---to their ability to affect a change in input code $X$, rather than in proportion to their actual prevalence in the field, which may be impossible to estimate except anecdotally.  Ultimately, some syntactic perturbations may not appear at all for any paragraphs if their target features (Section~\ref{ssec:methods_overview}) observationally are rare.  The sampling can be modified to up-weight such `rare' perturbations by attempting them before any other methods.

Because programs map uniquely to paragraphs, original paragraph $X_i$ is expanded into $n(i)=R_r+R_s$ variants $(\splitatcommas{X_{i,1},\dots,X_{i,R_s}, X_{i,R_s+1},\dots,X_{i,R_s+R_r+1}})$ input paragraphs extracted from the programs, which, together with $X_i$, form its group block (Section~\ref{ssec:measuring_robustness}).  

The user specifies an input $R$ that is the average number of \textit{paragraphs} (i.e., times) each perturbation method should appear in in the expanded dataset.  The \textit{program} multipliers $R_r$ and $R_s$ are determined to at least satisfy the desired $R$.  For instance, for \textit{genapp21} we set $R=5$, which gives $R_r=2$ and $R_s=7$ renaming and syntactic variants for each program, for a total of 90 programs (including unperturbed) and 210 paragraphs, ten times the size of the original.  A corpus of more programs with the same $R$ would be expanded by a lower factor (at least 3, given the constraints).  The goal is for the expanded benchmark to have a balanced representation of the perturbation methods, while avoiding full factorial expansion of each paragraph with each method that can perturb it.

Our utility allows for generation of layered paragraph- or program-based perturbation.  For instance, we can create a variant $\Delta_3(\Delta_2(\Delta_1(X_i)))$ with differing methods $\Delta_1\ne\Delta_2\ne\Delta_3$, and at each layer each method affects a change in its input.  
However, with multiple layers, we only see the result of the final cumulative perturbation, and some layers may even undo the changes of an earlier layer, making it difficult to measure the the effect of each method on the final any-metric-changed measure $\delta_{i,j}$. For these reasons we restrict our analysis to single-layer perturbation variants.  

A pipeline for a system that is program- rather than paragraph-based can be constructed similarly, by applying a different perturbation method to the entire program for each variant.  This is simpler than the paragraph-based system because the perturbed code unit and the analysis unit are both programs, so we don't have to deal with extracting paragraphs for analysis.

\subsection{Evaluation metrics
\label{ssec:dashboard_evaluation_metrics}
}

For this evaluation, we use only the following metrics $\mu_j, \:j=1,\dots,M$ (Section~\ref{ssec:metric_notation}) to detect a change in ($\delta_{i,j}\ne 1)$ (Section~\ref{ssec:measuring_robustness}); for simplicity, they are all boolean-valued, but one could also use numeric-valued metrics and possibly redefine $\delta_{i,j}$ to detect a change in $\mu_j$ above a threshold.  See \cite{froimovich2025quality} for details about these metrics.

\begin{itemize}[leftmargin=10px]  
    \item \textbf{parsable}: True if the Java translation ($f$) is parsable.
    \item \textbf{translated}: False if there was an internal error during the Java translation.  If False, the other metric values are missing.
    \item \textbf{not\_empty}: True if the translated Java code is non-empty.
    \item \textbf{non\_empty\_exec}: True if the Java code contains at least one executable statement.
    \item \textbf{not\_eq\_sig}: True if the COBOL procedure calls and Java methods do not have the same signatures.
    \item \textbf{proc\_invoke\_pass}: True if each method in the Java code has the appropriate parameters, given the corresponding COBOL.
    \item \textbf{var\_acc\_pass}: True if each variable access (read/write) in COBOL has a corresponding access of the same type in Java.
    \item \textbf{sql}: True if each COBOL SQL call was translated into the correct Java method.
    \item \textbf{file}: True if each COBOL file access call was translated into the correct Java method.
    \item \textbf{compilation\_passed}: True if the translated code after injection to the classes passes compilation.  Compilation can fail even if the translation is correct, for instance, if the classes are wrong.
    \item \textbf{uninjected\_compilation\_passed}: True if the Java classes \textit{without} the generated method pass compilation. 
    \item \textbf{laj\_pass}: True if the general quality LaaJ score (1--7) of the translation is 5 (`mostly accurate translation') or better.
\end{itemize}

\subsection{Evaluation dashboards
\label{ssec:dashboards}
}

We now show Grafana dashboards for monitoring the robustness measures and for manual examination.  All, except the debug one, allow comparison of results across multiple testsets and models.  We do not claim the results generalize to other models or datasets.


\subsubsection{Debug dashboard
\label{sssec:debug_dashboard}
}

The debug dashboard (Section~\ref{ssec:debug_dashboard} Figure~\ref{fig:debug_dashboard}) allows for manual inspection of results for one group block (Table~\ref{tab:perturbed_table}), i.e., an input paragraph $X_i$ and all its variant expansions.

In the top panel, each row is a paragraph variant, with the first being the original.  The leftmost boolean column (`any\_checker\_changed') represents the `$\delta$ (any)' column of whether any of the metrics changed relative to the original.  The other boolean columns contain the actual values $\mu_1(f),\mu_2(f),\dots$ of the metrics.  The boolean columns are colored by the corresponding values of the change indicators $\delta_1,\delta_2,\dots$ or $\delta$ (as in Table~\ref{tab:perturbed_table}), with red indicating there was a change.  Thus the cell `\textcolor{red}{true}' for metric var\_acc\_pass ($\mu_7$) on the eighth variant $j=8$ means that $\mu_7(\Delta_{i,8}(X_i))=1$ (True) and $\delta_{i,8,7}=1$ (there was a change), since $\mu_7(X_i)=0$ (False).

The middle panels show partial views of the input COBOL code $X_i$ and its perturbed variants $\Delta_{i,j}(X_i),\: j=1,\dots,n(i)$.  The highlighted portions show the differences relative to the original, that is, the code perturbations.  The bottom panels show the generated Java translations $f(X_i)$ and $f(\Delta_{i,j}(X_i))$, also with highlights.

We emphasize that the nature of LLM-based generative systems is that direct cause-and-effect of perturbations may be difficult to ascertain, or may not even exist.  We may not be able to say that, say, `changing tokens to uppercase in COBOL tends to cause \_\_\_ change in the Java translation'.  Even if such relationships exist, it would likely take significant manual inspection to deduce; furthermore, each different LLM used in a system could have different effects, which would multiply the effort if we are to compare different LLMs.  Thus, we suggest using easily-measured criteria like changes in the metric values to assess and compare robustness.

\subsubsection{Overall robustness scores
\label{sssec:overall_dashboard}
}

\begin{figure}[h]
  \centering
    \includegraphics[width=\linewidth]{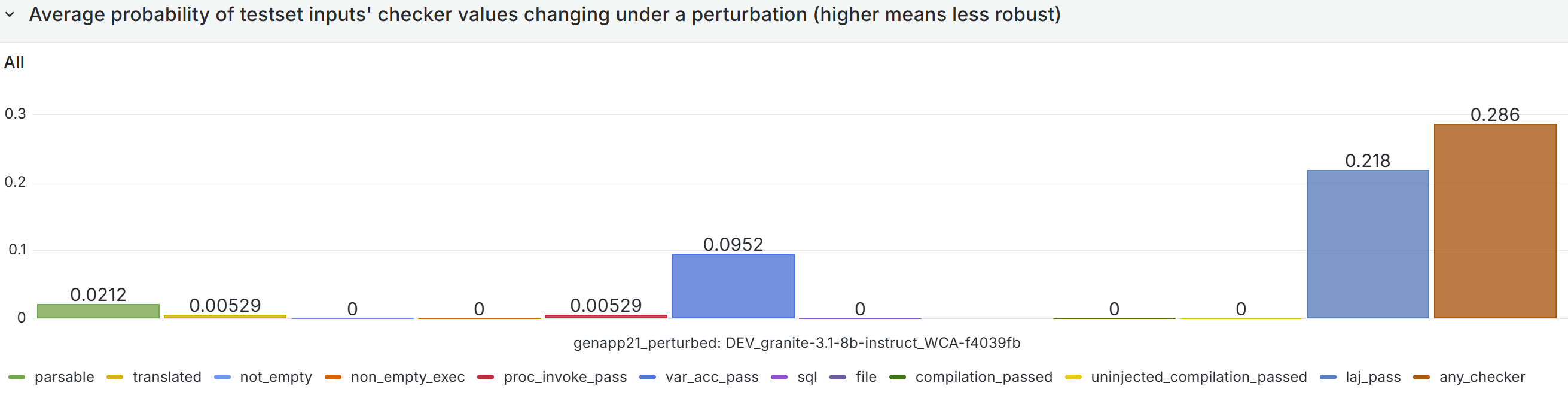}
    \includegraphics[width=\linewidth]{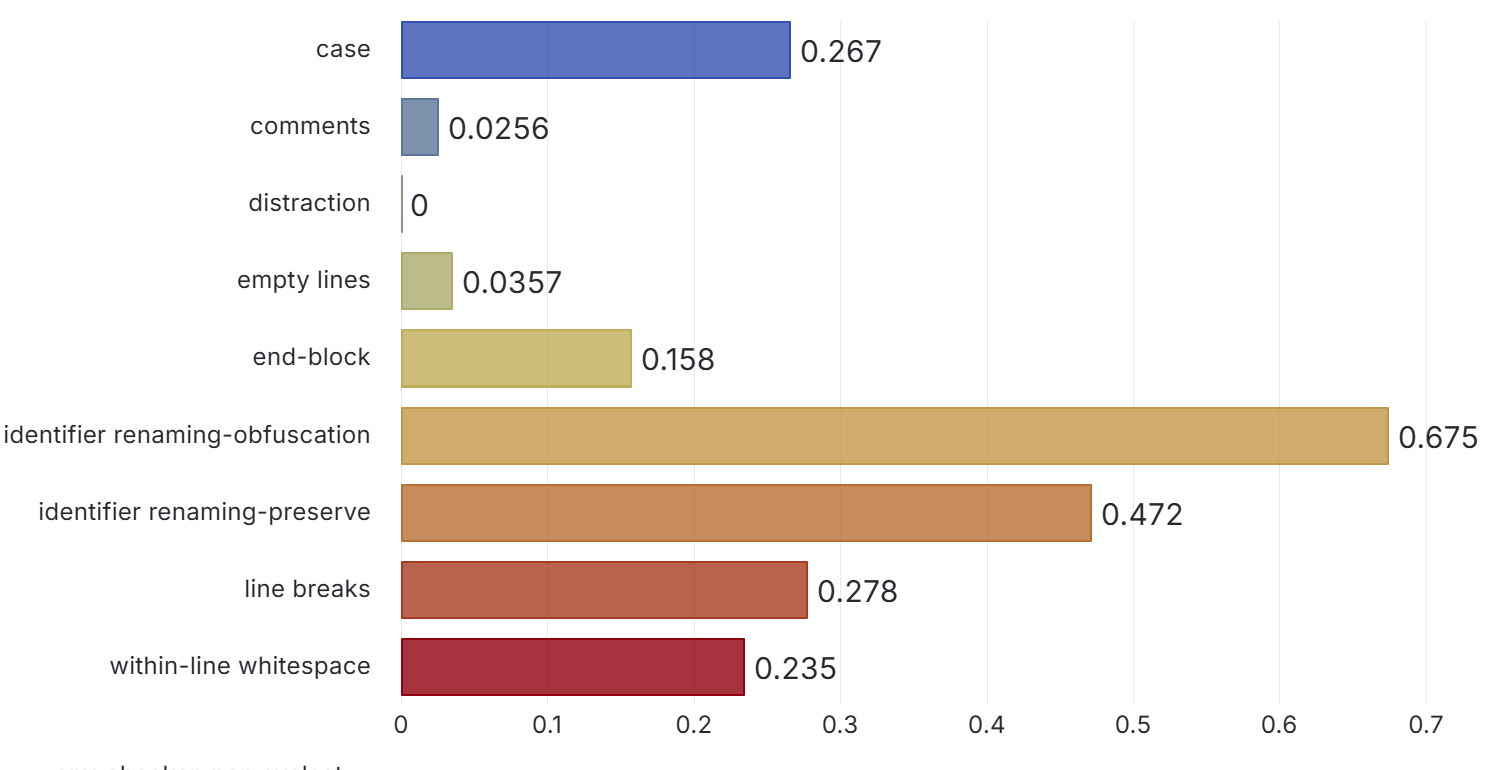}
    \includegraphics[width=\linewidth]{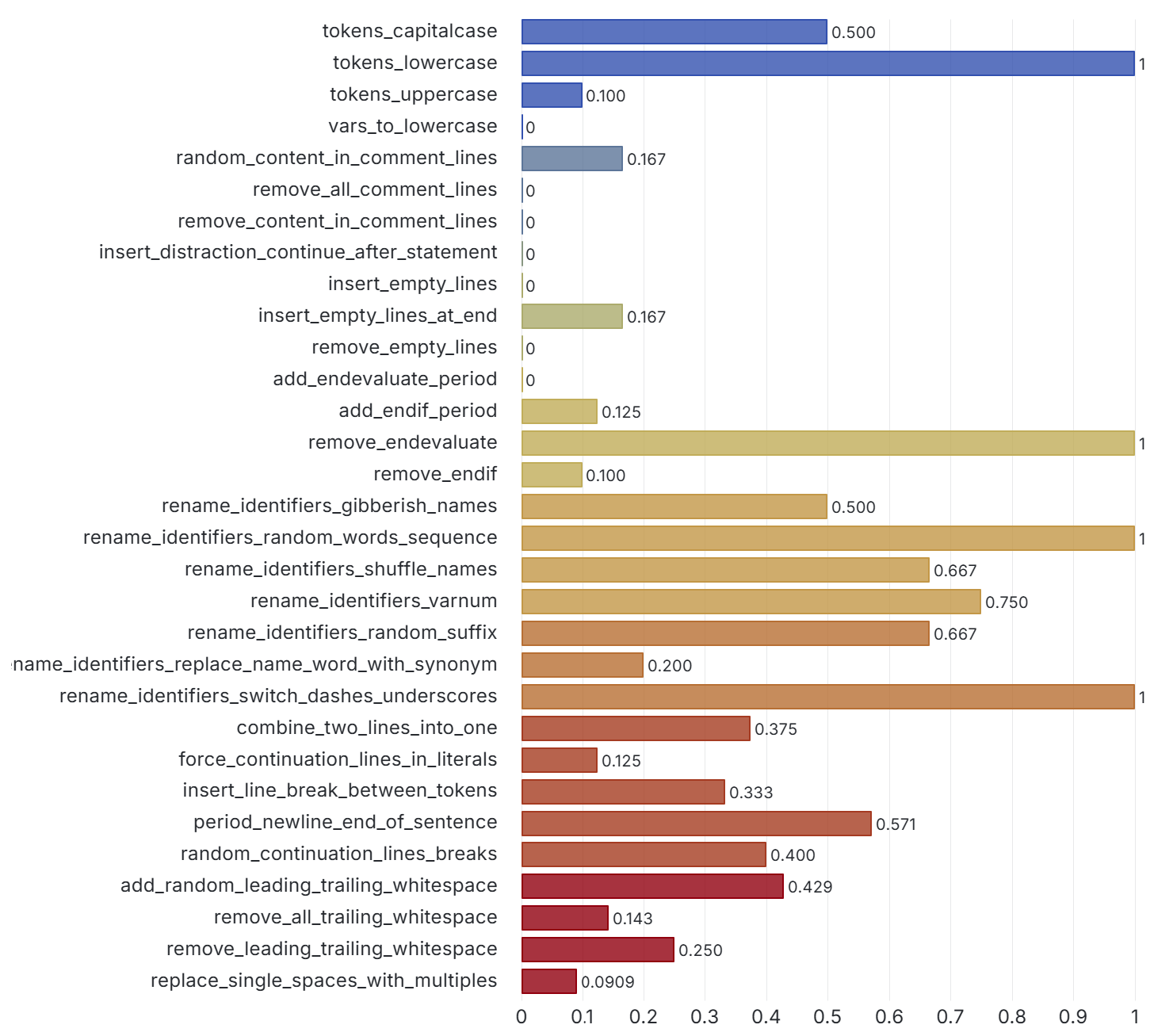}

  \caption{\label{fig:overall_dashboard}  Top-line statistical summary.\\
  Top: Average probability of any input changing a metric $\mu$ (checker).\\
  Middle, bottom: Average category- ($D$) and perturbation method-conditional probability of changing any metric ($\bd_D$).
  }
\end{figure}

Figure~\ref{fig:overall_dashboard} displays the summary statistics for the expanded dataset.  The top panel shows $\bd_m, \:m=1,\dots,M$, the average probability of a random perturbation changing the value of each of our $M$ evaluation metrics.  The rightmost column is $\bd$, the average probability of at least one of the metrics changing, which is our proposed anti-robustness metric.  Here, given the distributions of our perturbation types and input paragraphs, this probability is about 29\%.

The second and third panels show the $\bd_D$, the average perturbation-method-conditional probability of changing any of the metrics, for the categories and individual perturbations, respectively.  In the third panel, each method is colored according to its category from the second panel.  These panels indicate that the two identifier-renaming categories tended to be most likely to change metric values, that is, the generation pipeline was least robust to these types of changes.  In particular, the `obfuscation' renaming type, which alters more of the information inherent in the identifier names that does the `preserve' type, had the larger effect of the two.

Of the syntactic categories, changing the within-line whitespace, line breaks, and token case had the strongest effect, though these were lower than the renaming categories.  It is likely, on the basis of these observations, that incorporating pre-processing steps to standardize the input could improve the robustness of the pipeline.  For instance, it is fairly easy to standardize token case to uppercase, between-token space to single spaces, remove empty lines, etc., and employ SFX form; but the logic may not extend to undoing user-specified between-token line breaks, which can be informative.


\begin{figure*}[h]
  \centering
  \includegraphics[width=0.95\textwidth]{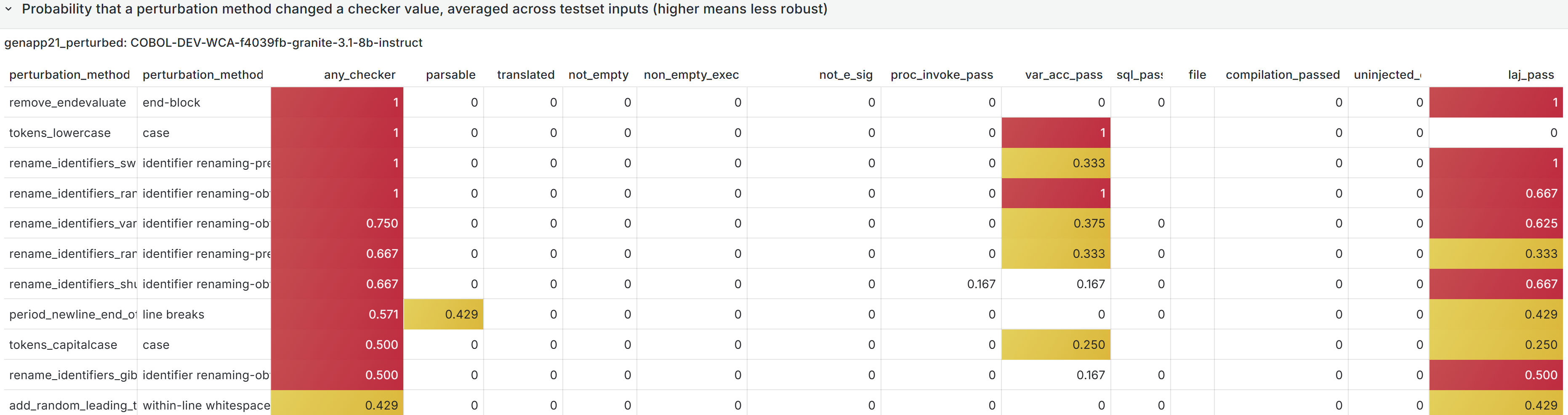}
  \includegraphics[width=0.95\textwidth]{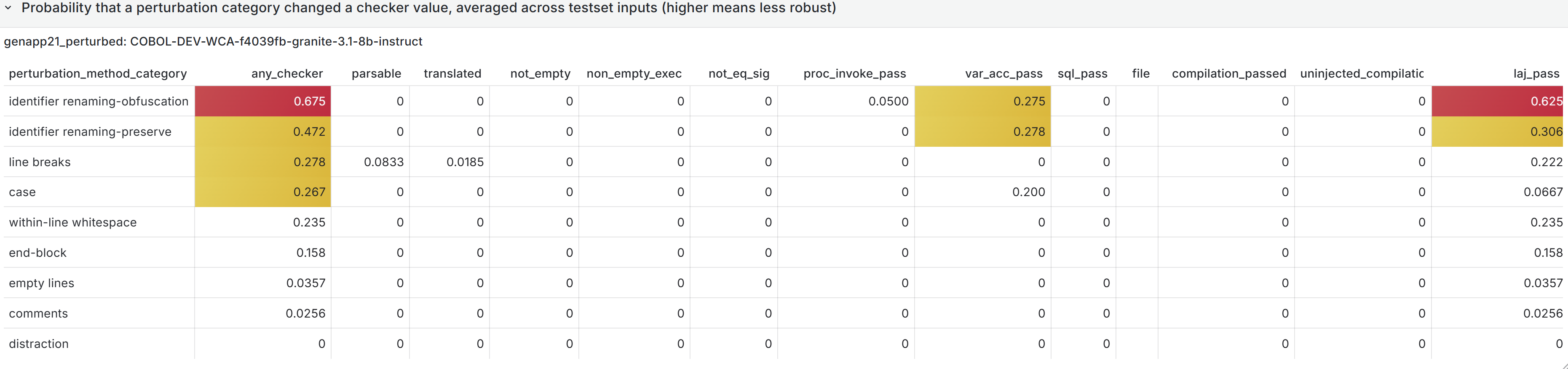}
  \caption{\label{fig:overall_dashboard2}
  By-method (top) and category (bottom) probabilities of each evaluation metric changing value.
  }
\end{figure*}

Figure~\ref{fig:overall_dashboard2} shows the individual perturbation method and category ($D$)-conditional probabilities of changing any metric value ($\bd_D$, shown in the second and third panel of Figure~\ref{fig:overall_dashboard}) and each metric $m$ value on its own ($\bd_{D,m}$).  See Section~\ref{ssec:robustness_proxy_measures} for calculations.  Change probabilities between [0.25, 0.5] and above 0.5 are indicated by yellow and red background color.  We can see that the var\_acc\_pass and laj\_pass metrics were the most sensitive to perturbations.

We typically expect $X_i$ to be in `ideal' form and that any metric change by a perturbation is for the worse.  However, this is not always the case.  For instance, period\_newline\_end\_of\_sent is likely to change parsability  (Figure~\ref{fig:overall_dashboard2}), but this is actually mostly 
a positive change of the variants becoming parsable when the original was not.  However, this does not mean that the (parsable) variant translation $f(\Delta_{i,j}(X_i))$ is strictly better than $f(X_i)$ of the original; $f(\Delta_{i,j}(X_i))$ could still be very incorrect but trivially parsable.  The metric still serves as an indicator of change, and hence non-robustness.

\subsubsection{Input-specific robustness scores
\label{sssec:sample_dashboard}
}

\begin{figure*}
  \centering    \includegraphics[width=0.95\textwidth]{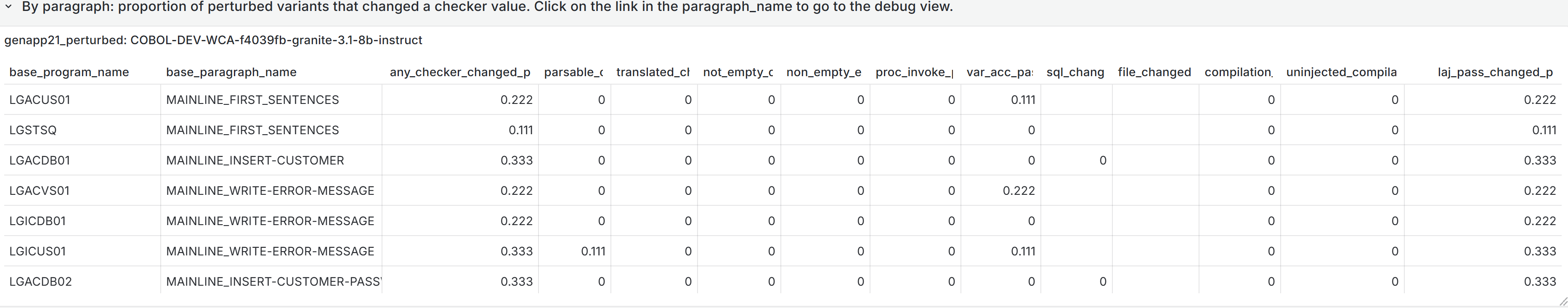}

  \caption{\label{fig:sample_dashboard}  Input-level statistical summary: Average probability of each/any metric changing, on perturbations of each input $X_i$.}
\end{figure*}

Figure~\ref{fig:sample_dashboard} shows summary robustness statistics for each input $X_i$ in the original corpus.  As shown in Section~\ref{ssec:robustness_proxy_measures}, the per-input rate of change of metric $m$, denoted $\bd_{i,\cdot,m}$, is the probability this metric value changed, across the perturbed variants of $X_i$.  The input-$i$ `any metric changed' probability is $\bd_i=\frac{\sum_{j=1}^{n(i)}\delta_{i,j}}{n(i)}$, the average of the `any change' indicator $\delta_{i,j}$ for its variants $j$.  This is the first numeric column in the table.  

Recall, we subject each input $X_i$ to $n(i)$ \textit{different} perturbation methods $D$, which are a random selection from the subset of the available methods $\mathcal{D}$ that can perturb $X_i$.  The inputs $X_i$ are not themselves 'units' of interest for robustness analysis, but are only useful as a collection to represent some sample distribution over which we will assess the robustness of our system; this assumes that collected sample is in fact representative of the of inputs in the field, which is the population distribution of interest.  Thus, we suggest the input-level tables be used simply for diagnosing the contributing inputs $X_i$ to the overall dataset robustness score $\bd$, and not as `robustness' measures of the inputs $X_i$ themselves.

%% file: sections/conclusion.tex
\section{Conclusion \label{sec:conclusion}}

We introduced a framework to evaluate the perturbation robustness of a generative system receiving COBOL code inputs.  The particular setting is COBOL-to-Java translation, but the same framework and perturbations are relevant to other tasks, such as natural language code summarization or COBOL code completion.  The strategy we employ, of using different meaning-preserving variations of a given input, has been commonly-used in the literature on robustness of code and natural language generation.  We assess the robustness---the degree to which the system outputs change due to different perturbed inputs---by observing the rate of value change of evaluation metrics that measure specific aspects of the generative output; these metrics help ease the task of detecting problematic instances of non-robustness.  

In particular, we discussed the effect of perturbation of continuation lines in fixed-form output, which is of specific interest in COBOL.  Lastly, we showed the use of dynamic dashboards that assist with monitoring the overall robustness of a system with a particular LLM on a representative benchmark, as well as debugging particular instances.  However, since straightforward causal effects of perturbations to LLM inputs can be difficult to formulate, we recommend using the top-line robustness monitoring measures rather than inspecting the outputs of individual inputs. 

%% file: sections/appendix.tex
\appendix
\onecolumn

\section{Appendix
\label{sec:appendix}
}

\subsection{\label{ssec:perturbation_definitions}}

In Table~\ref{tab:method_examples} we give definitions and examples of each of the perturbation methods.

\needspace{5cm}
\begin{longtable}{|p{2cm}|p{2cm}|p{2.5cm}|p{5cm}|p{5cm}|}
    \hline
    \textbf{Method} & \textbf{Category} & \textbf{Description} &
    \textbf{Original Input} &
    \textbf{Perturbed output}
    \\
    \hline
    
    add\_endevaluate\-\_period &
	end-block &
	If possible (when it is in a sentence but not the last statement), replace END-EVALUATE not having a period with a period. &         
        \raisebox{-1\height}{\includegraphics[width=5cm]{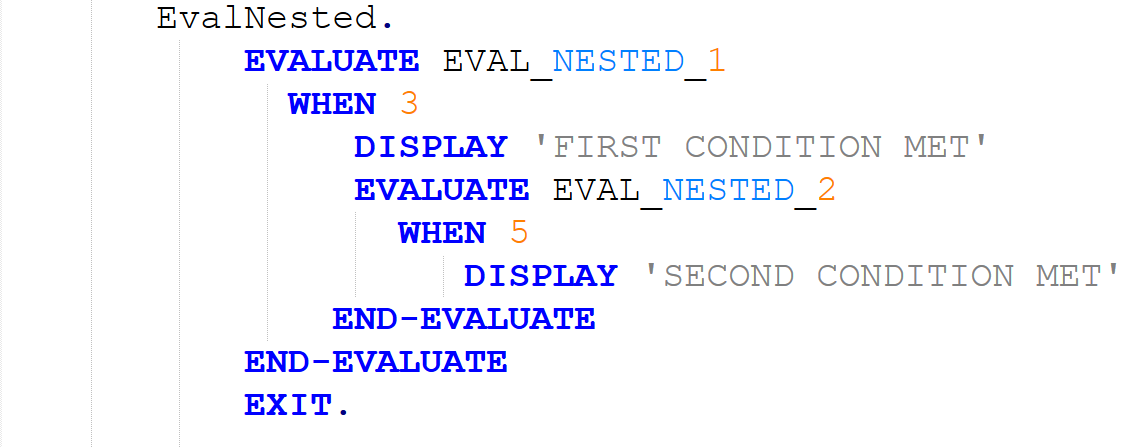}} &

        \raisebox{-1\height}{\includegraphics[width=5cm]{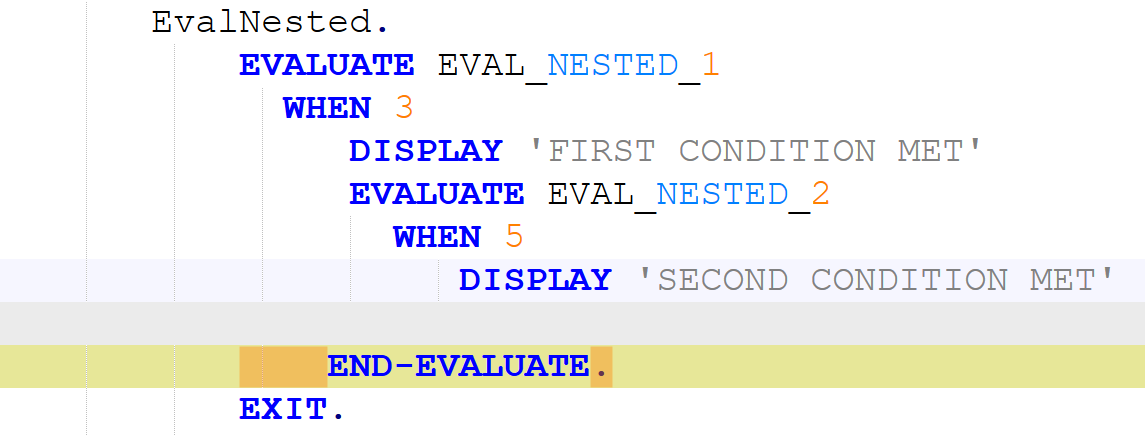}} \\
        \hline

    \hline

    add\_endif\-\_period &
	end-block &
	If possible (when it is in a sentence but not the last statement), replace END-IF not having a period with a period. &
        \raisebox{-1\height}{\includegraphics[width=5cm]{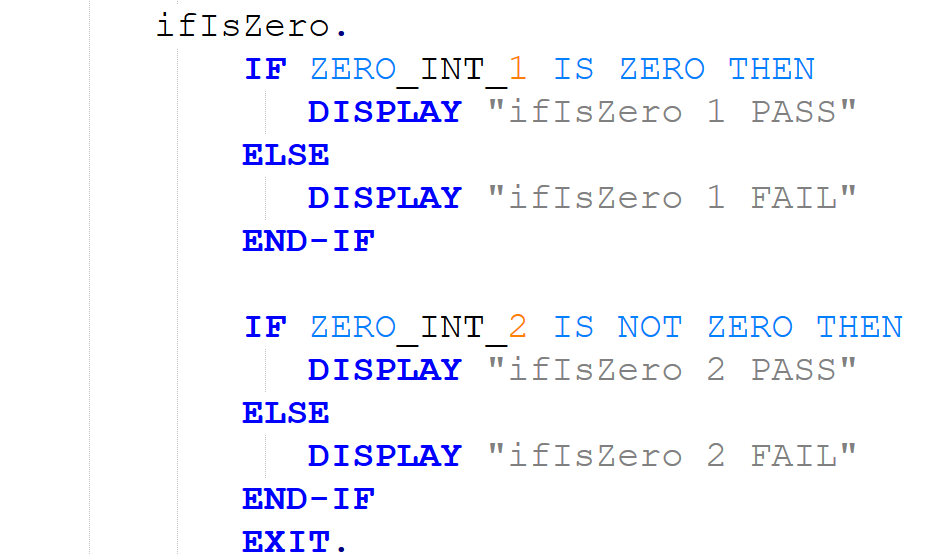}} &

        \raisebox{-1\height}{\includegraphics[width=5cm]{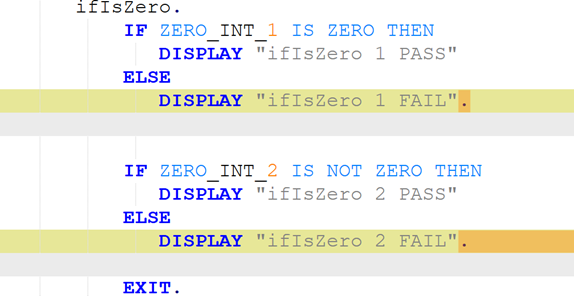}} \\
        \hline

    add\_endstring\-\_period &
	end-block &
	If possible (when it is in a sentence but not the last statement), replace END-STATEMENT not having a period with a period. & & \\
    \hline

    add\_random\-\_leading\_trailing\-\_whitespace &
	within-line whitespace &
	Add random amounts of leading and trailing whitespace (if non-empty and non-comment lines) to each line, retaining the line’s belonging in area A or B. &
        \raisebox{-1\height}{\includegraphics[width=5cm]{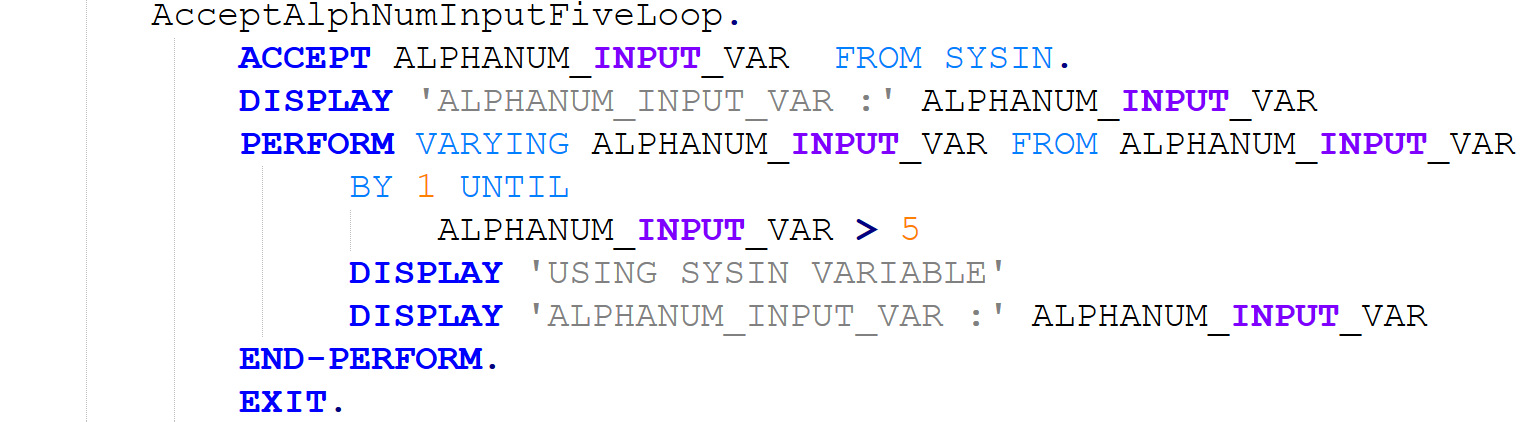}} &

        \raisebox{-1\height}{\includegraphics[width=5cm]{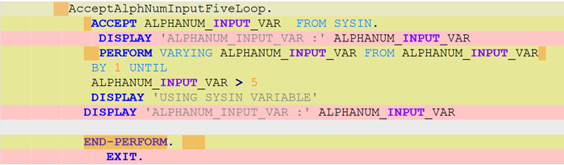}} \\
        \hline

    combine\_two\-\_lines\_into\_one &
	line breaks &
	Combine successive lines into one, if the length would not exceed 72.  The two lines must both be non-empty, not comment lines, and not part of a continuation block; also, the second may not be part of Area A. If these conditions are met, the merging is performed with a coin toss of probability 0.7. &
            \raisebox{-1\height}{\includegraphics[width=5cm]{figures/acceptalphnum.png}} &

        \raisebox{-1\height}{\includegraphics[width=5cm]{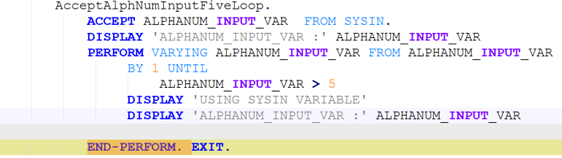}} \\
        \hline

    force\-\_continuation\-\_lines\_in\-\_literals &
	line breaks &
	If possible, force each literal to be split in a random location with a continuation line.  \textbf{Only valid for fixed-form output}. &
                \raisebox{-1\height}{\includegraphics[width=5cm]{figures/acceptalphnum.png}} &

        \raisebox{-1\height}{\includegraphics[width=5cm]{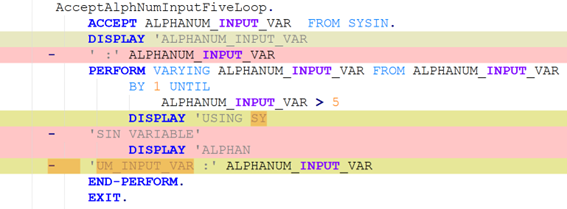}} \\
    \hline

    insert\-\_distraction\-\_continue\_after\-\_statement &
	distraction &
	With probability 0.3, insert a CONTINUE statement on the line after a statement. &
                    \raisebox{-1\height}{\includegraphics[width=5cm]{figures/acceptalphnum.png}} &

        \raisebox{-1\height}{\includegraphics[width=5cm]{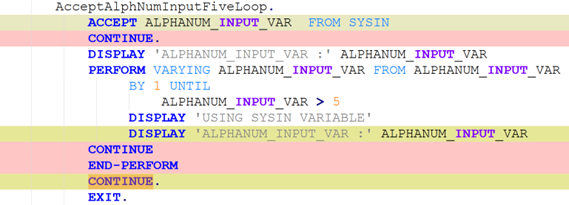}} \\
    \hline

    insert\_empty\-\_lines &
	empty lines &
	With probability 0.5, insert an empty line after a line (as long as the line does not have a continuation mark at the beginning). &
        \raisebox{-1\height}{\includegraphics[width=5cm]{figures/acceptalphnum.png}} &

        \raisebox{-1\height}{\includegraphics[width=5cm]{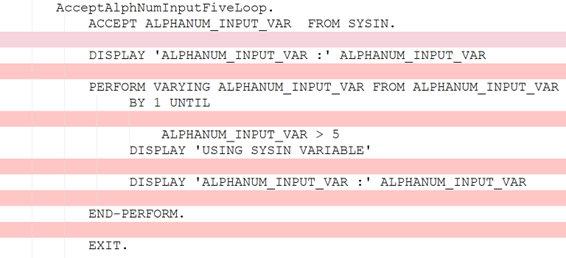}} \\
    \hline

    insert\_empty\-\_lines\_at\_end &
	empty lines &
	Insert four empty lines at the end of input. &
            \raisebox{-1\height}{\includegraphics[width=5cm]{figures/acceptalphnum.png}} &

        \raisebox{-1\height}{\includegraphics[width=5cm]{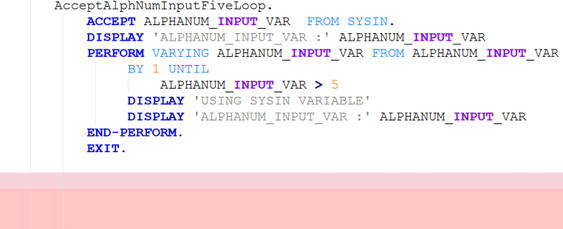}} \\
    \hline
    insert\_line\-\_break\_between\-\_tokens &
	line breaks &
	With coin toss probability 0.2, insert new line break between two successive tokens. &
            \raisebox{-1\height}{\includegraphics[width=5cm]{figures/acceptalphnum.png}} &

        \raisebox{-1\height}{\includegraphics[width=5cm]{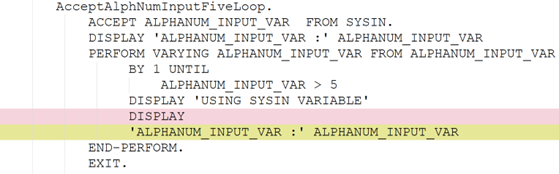}} \\
    \hline

    period\-\_newline\_end\-\_of\_sentence &
	line breaks &
	Insert a line break before the period at the end of a sentence. &
            \raisebox{-1\height}{\includegraphics[width=5cm]{figures/acceptalphnum.png}} &

        \raisebox{-1\height}{\includegraphics[width=5cm]{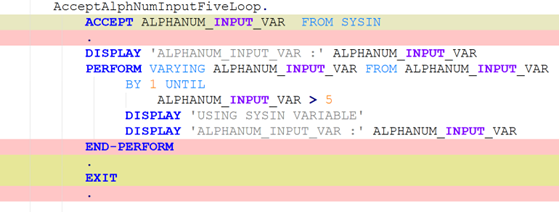}} \\
    \hline

    random\_content\-\_in\_comment\-\_lines &
	comments &
	Replace the content of any comments with a sequence of randomly selected words from a fixed list, filling the available space up to column 72. &
        \raisebox{-1\height}{\includegraphics[width=5cm]{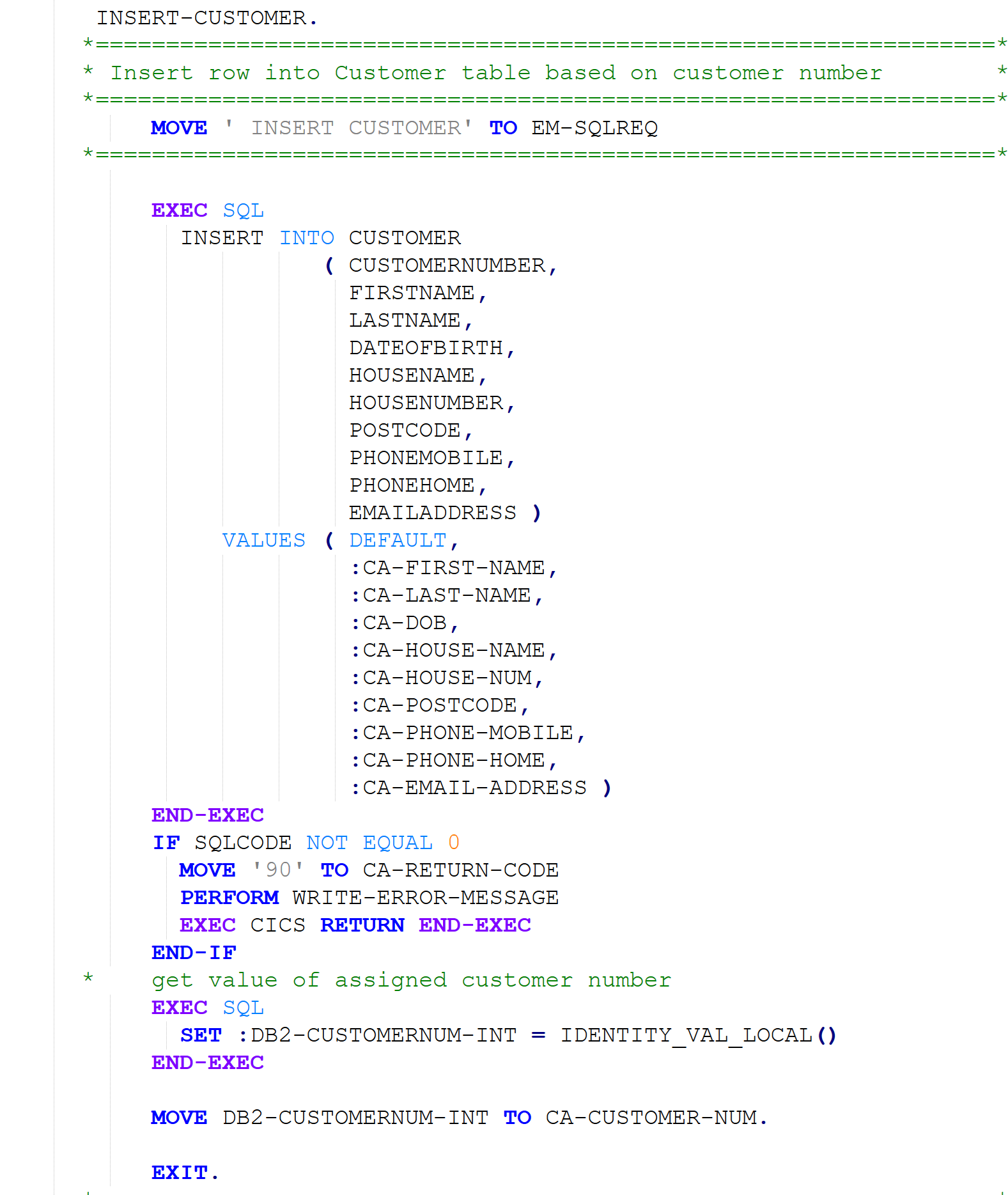}} &

        \raisebox{-1\height}{\includegraphics[width=5cm]{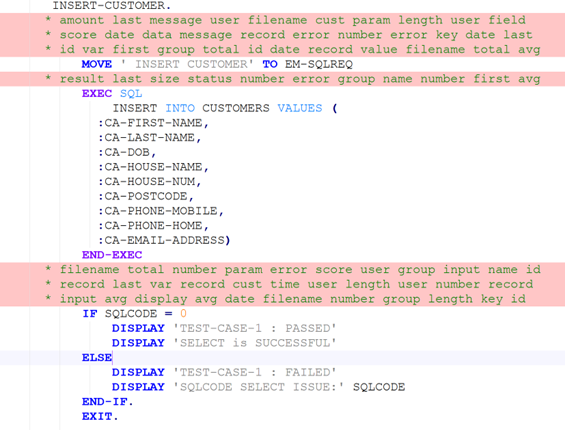}} \\
        \hline

    random\-\_continuation\-\_lines\_breaks	 & line breaks &	Insert a continuation line in a line that is longer than 20 characters, with probability 0.5, at a random legal location. &

            \raisebox{-1\height}{\includegraphics[width=5cm]{figures/acceptalphnum.png}} &

        \raisebox{-1\height}{\includegraphics[width=5cm]{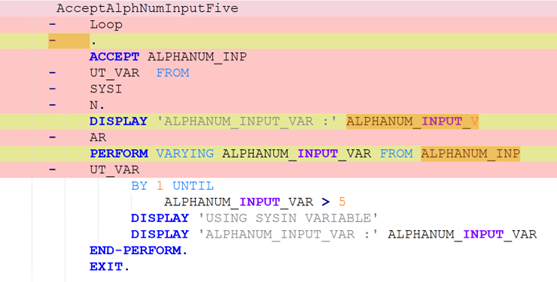}} \\
        \hline

    remove\_all\-\_comment\_lines &
	comments &	Entirely remove any lines containing comments. &
            \raisebox{-1\height}{\includegraphics[width=5cm]{figures/insertcustomer.png}} &

        \raisebox{-1\height}{\includegraphics[width=5cm]{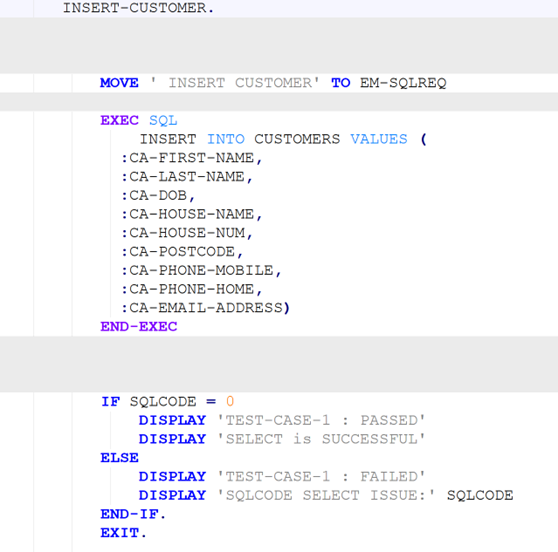}} \\
        \hline

    remove\_all\-\_trailing\-\_whitespace &
	within-line whitespace	& Remove all  trailing whitespace in lines (if there was whitespace, leaves a single space in case the next line continues it). &
            \raisebox{-1\height}{\includegraphics[width=5cm]{figures/acceptalphnum.png}} &

        \raisebox{-1\height}{\includegraphics[width=5cm]{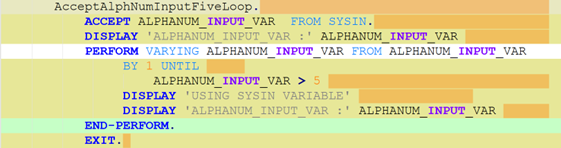}} \\
        \hline

    remove\_content\-\_in\_comment\-\_lines &
	comments &	For each comment line, remove the content after the comment indicator, but leave the comment indicator.  &
            \raisebox{-1\height}{\includegraphics[width=5cm]{figures/insertcustomer.png}} &

        \raisebox{-1\height}{\includegraphics[width=5cm]{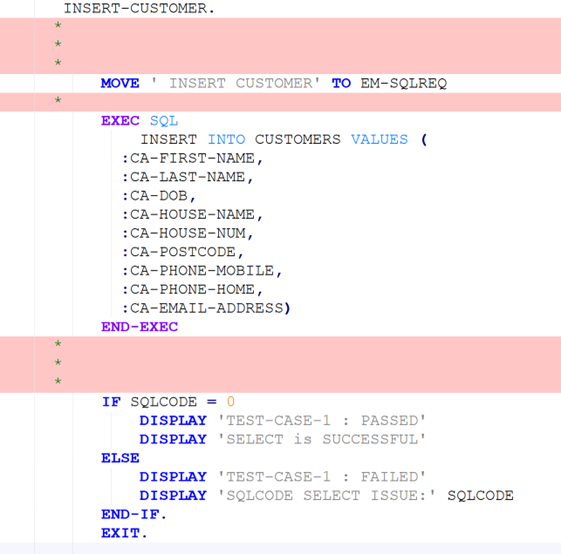}} \\
        \hline

    remove\_empty\-\_lines &
	empty lines &
	Remove all empty lines. &
            \raisebox{-1\height}{\includegraphics[width=5cm]{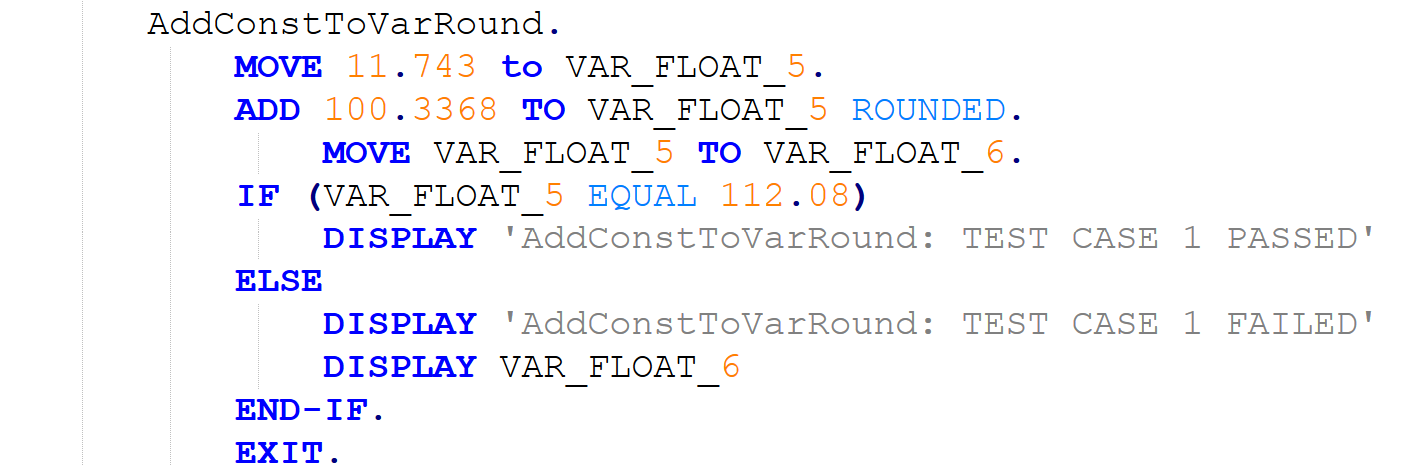}} &

        \raisebox{-1\height}{\includegraphics[width=5cm]{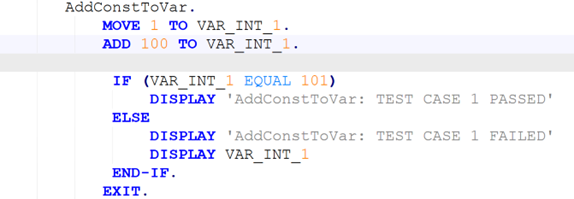}} \\
        \hline

    remove\-\_endevaluate &
	end-block &
	If possible (when it is the last statement in a sentence), remove END-EVALUATE, leaving the period after it. &
            \raisebox{-1\height}{\includegraphics[width=5cm]{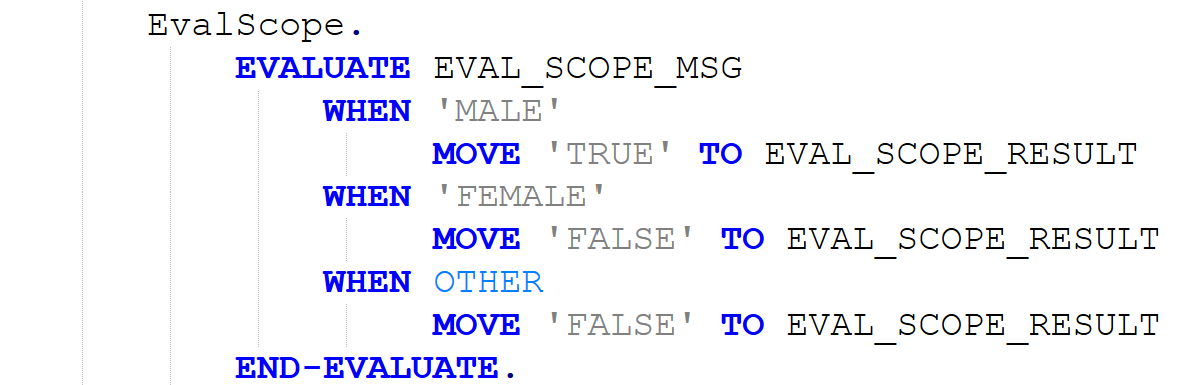}} &

        \raisebox{-1\height}{\includegraphics[width=5cm]{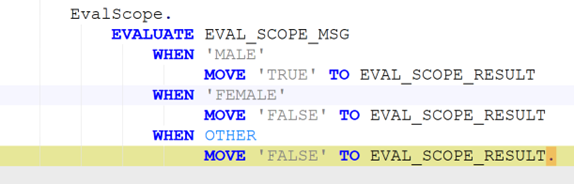}} \\
        \hline

    remove\-\_endif &
	end-block &
	If possible (when it is the last statement in a sentence), remove END-IF, leaving the period after it. &
            \raisebox{-1\height}{\includegraphics[width=5cm]{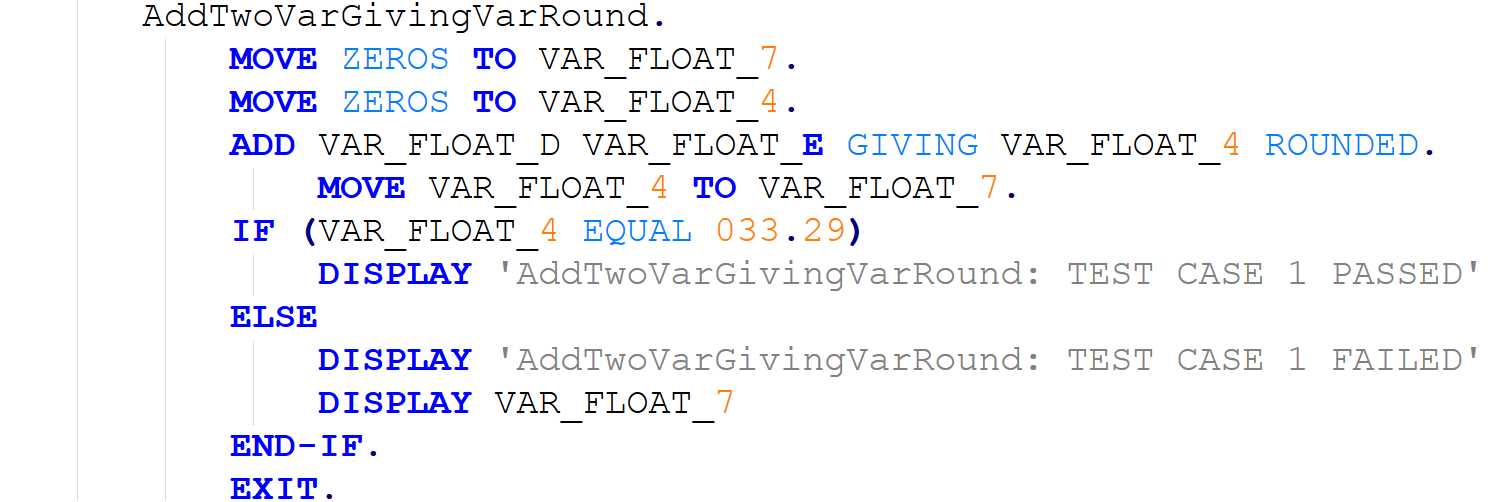}} &

        \raisebox{-1\height}{\includegraphics[width=5cm]{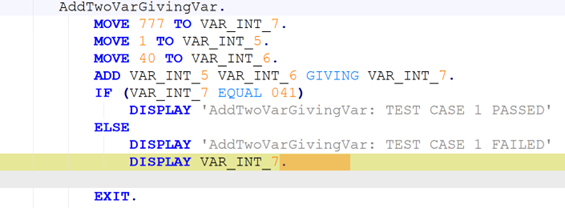}} \\
        \hline

    remove\-\_endstring &
	end-block &
	If possible (when it is the last statement in a sentence), remove END-STRING, leaving the period after it.
    & & \\
    \hline

    remove\_leading\-\_trailing\-\_whitespace &
	within-line whitespace &
	Remove all trailing whitespace and trailing whitespace in lines, while retaining the area separation. Area A lines have 7 leading spaces, and Area B lines have 11. &
    \raisebox{-1\height}{\includegraphics[width=5cm]{figures/acceptalphnum.png}} &

        \raisebox{-1\height}{\includegraphics[width=5cm]{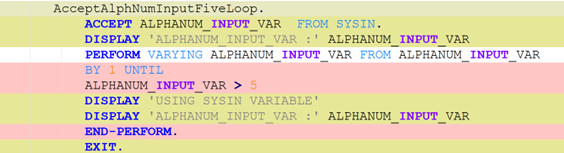}} \\
        \hline

    replace\_single\-\_spaces\_with\-\_multiples &
	within-line whitespace &
	Perturbation targets instances where single spaces separate tokens in a line.  With probability 0.5, replace a single space with 2 or more  spaces (the number is selected using a Poisson distribution).  By default, the expansion of the spaces is constrained by limiting the total length to 72. &
    \raisebox{-1\height}{\includegraphics[width=5cm]{figures/acceptalphnum.png}} &

        \raisebox{-1\height}{\includegraphics[width=5cm]{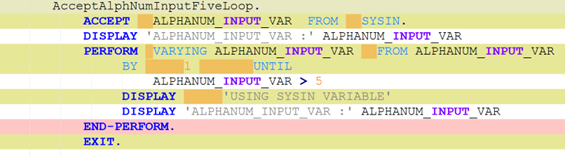}} \\
        \hline

    tokens\-\_capitalcase &
	case &	With probability 0.5, change the case of a (non-literal) token to capital case (first letter capitalized, others are lowercase). &
        \raisebox{-1\height}{\includegraphics[width=5cm]{figures/acceptalphnum.png}} &

        \raisebox{-1\height}{\includegraphics[width=5cm]{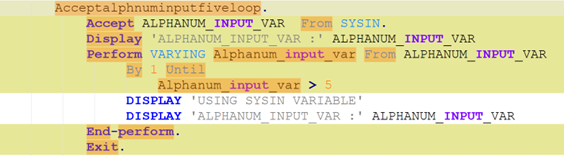}} \\
        \hline

    tokens\-\_lowercase &
	case &	With probability 0.5, change the case of a (non-literal) token to lower case. &
        \raisebox{-1\height}{\includegraphics[width=5cm]{figures/acceptalphnum.png}} &

        \raisebox{-1\height}{\includegraphics[width=5cm]{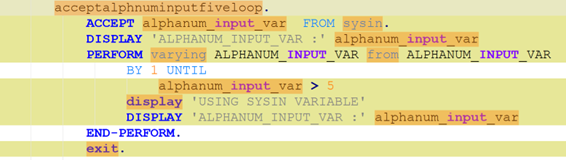}} \\
        \hline

    tokens\-\_uppercase &
	case &	With probability 0.5, change the case of a (non-literal) token to upper case. &
            \raisebox{-1\height}{\includegraphics[width=5cm]{figures/acceptalphnum.png}} &

        \raisebox{-1\height}{\includegraphics[width=5cm]{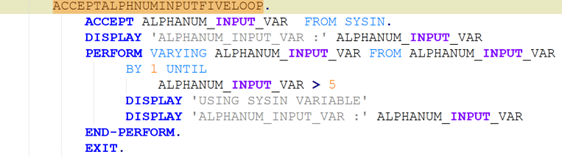}} \\
        \hline

    vars\_to\-\_lowercase &
	case &	With probability 0.5, change the case of an identifier to lowercase. &
                \raisebox{-1\height}{\includegraphics[width=5cm]{figures/acceptalphnum.png}} &

        \raisebox{-1\height}{\includegraphics[width=5cm]{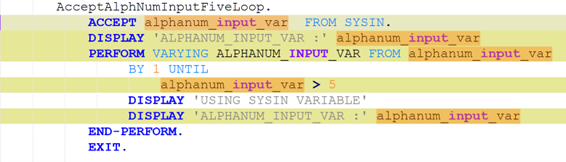}} \\
        \hline

    rename\-\_identifiers\-\_gibberish\-\_names &
	identifier renaming-obfuscation &
	Replace identifier names (if legal) with a random 8-letter string.  Replacements are done globally throughout the program using the same mapping. &

                \raisebox{-1\height}{\includegraphics[width=5cm]{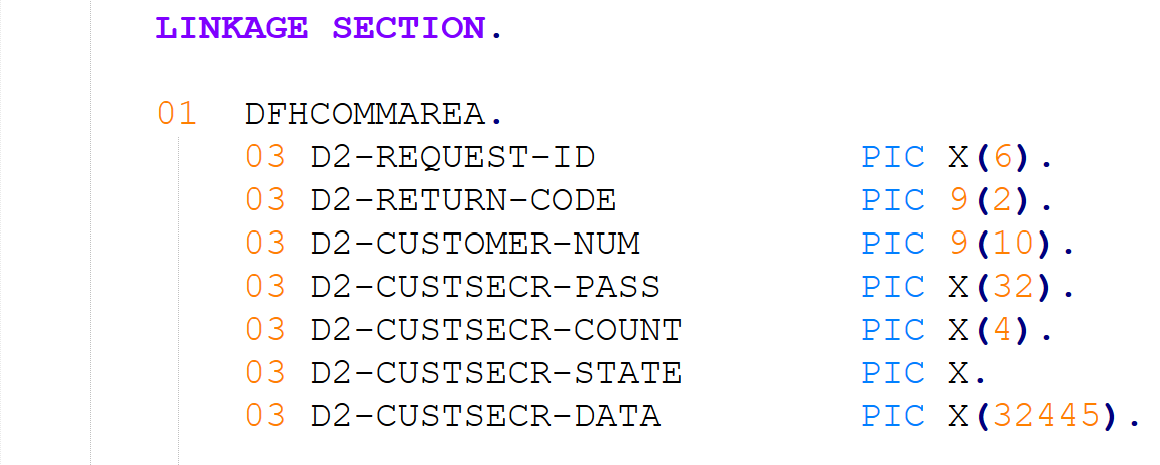}} &

        \raisebox{-1\height}{\includegraphics[width=5cm]{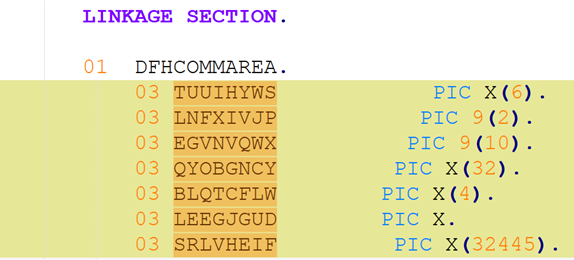}} \\
        \hline

    rename\-\_identifiers\-\_random\_suffix &
	identifier renaming-preserve &
	Append “-X” to each identifier name (if legal), where “X” is a random single letter. Replacements are done globally throughout the program using the same mapping. &
                \raisebox{-1\height}{\includegraphics[width=5cm]{figures/linkagesection.png}} &

        \raisebox{-1\height}{\includegraphics[width=5cm]{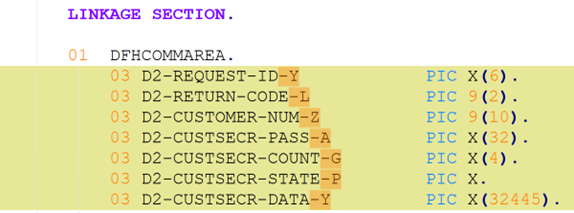}} \\
        \hline

    rename\-\_identifiers\-\_random\_words\-\_sequence &
	identifier renaming-obfuscation &
	Replace identifier names (if legal) with a sequence of 3 random words. Replacements are done globally throughout the program using the same mapping. &

                    \raisebox{-1\height}{\includegraphics[width=5cm]{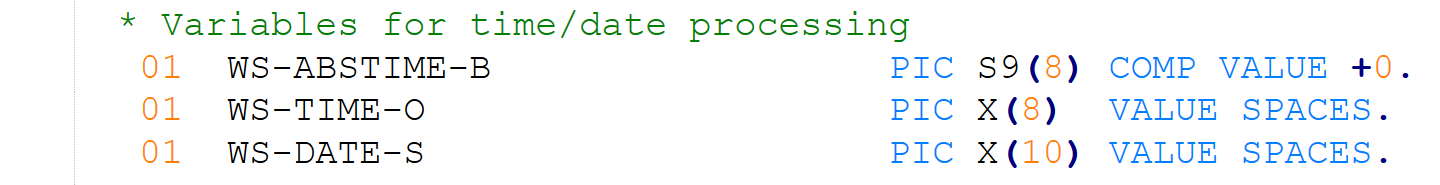}} &

        \raisebox{-1\height}{\includegraphics[width=5cm]{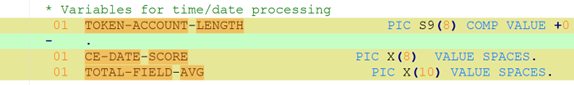}} \\
        \hline

    rename\-\_identifiers\-\_replace\_name\-\_word\_with\-\_synonym &
	identifier renaming-preserve &
	With probability 0.5 (if legal) replace an English word in an identifier name with a synonym. Replacements are done globally throughout the program using the same mapping. &

        \raisebox{-1\height}{\includegraphics[width=5cm]{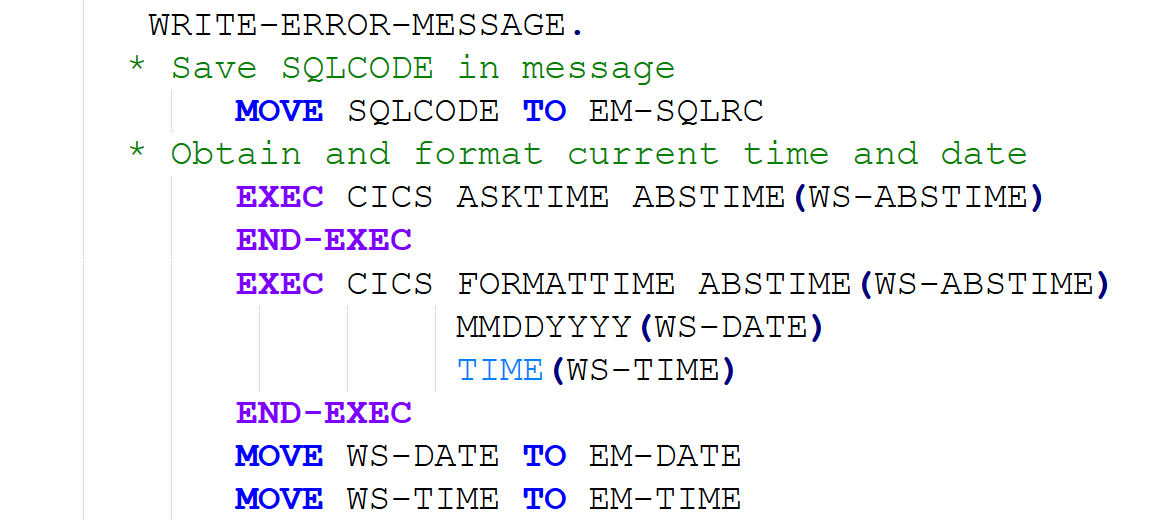}} &

        \raisebox{-1\height}{\includegraphics[width=5cm]{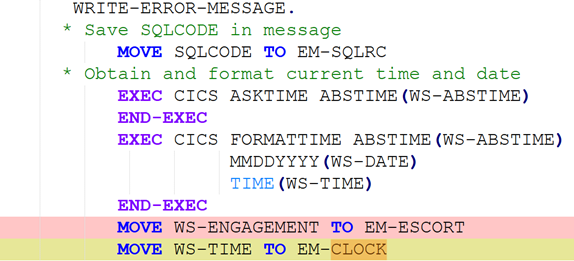}} \\
        \hline

    rename\-\_identifiers\-\_shuffle\_names &
	identifier renaming-obfuscation &
	Replace an identifier name with another random identifier from the same program. Replacements are done globally throughout the program using the same mapping. &
        \raisebox{-1\height}{\includegraphics[width=5cm]{figures/writeerrormessage.png}} &

        \raisebox{-1\height}{\includegraphics[width=5cm]{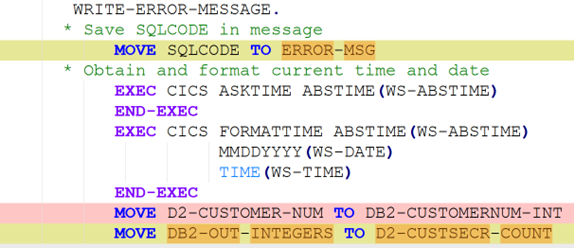}} \\
        \hline

    rename\-\_identifiers\-\_switch\_dashes\-\_underscores &
	identifier renaming-preserve &
	Replace dashes with underscores in an identifier name, and vice versa. Replacements are done globally throughout the program using the same mapping. &
            \raisebox{-1\height}{\includegraphics[width=5cm]{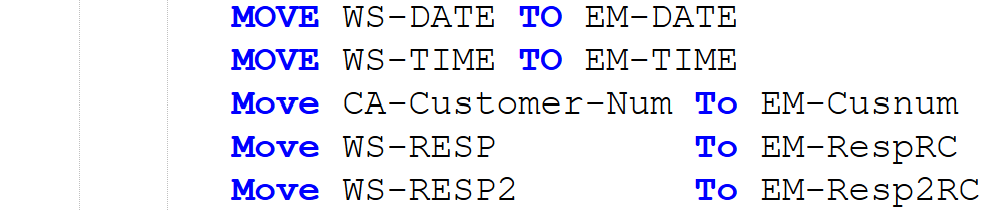}} &

        \raisebox{-1\height}{\includegraphics[width=5cm]{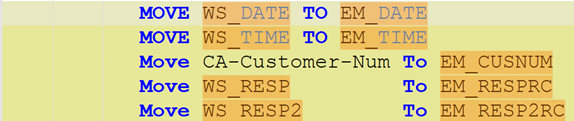}} \\
        \hline

    rename\-\_identifiers\-\_varnum &
	identifier renaming-obfuscation &
	Replace an identifier name with VAR-\#, where \# is a random integer. Replacements are done globally throughout the program using the same mapping. &
            \raisebox{-1\height}{\includegraphics[width=5cm]{figures/variablesfortime.png}} &

        \raisebox{-1\height}{\includegraphics[width=5cm]{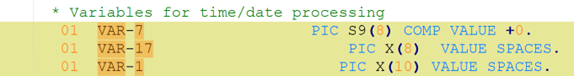}} \\
        \hline

    \caption{\label{tab:method_examples} Examples of input and output for perturbation methods.}
\end{longtable}

\subsection{Debug dashboard
\label{ssec:debug_dashboard}}

\begin{figure*}
  \centering
  \includegraphics[width=\textwidth]{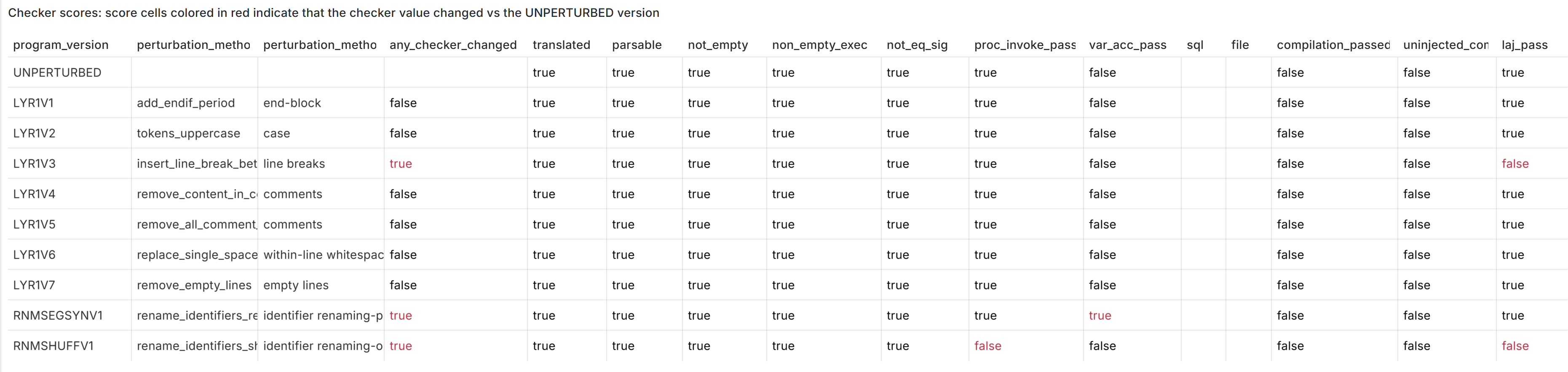}
  \includegraphics[width=\textwidth]{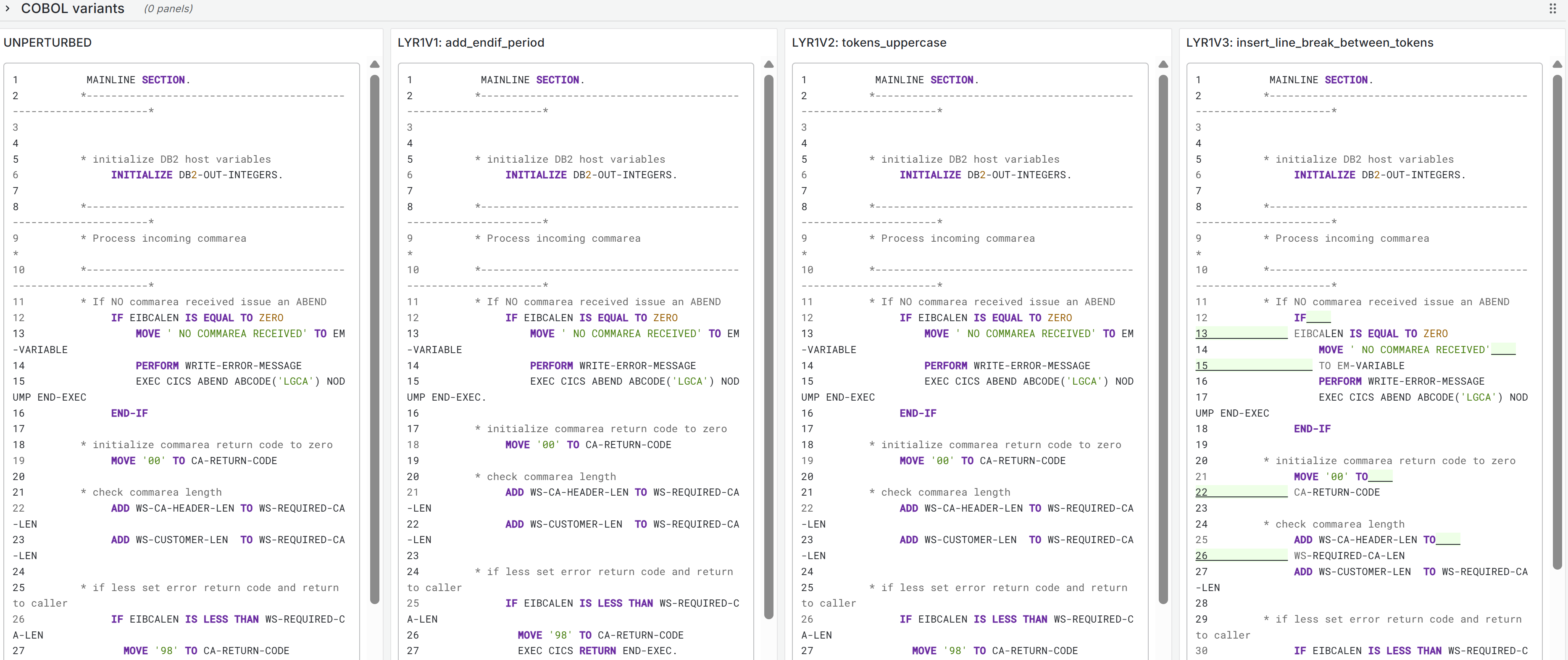}
  \includegraphics[width=\textwidth]{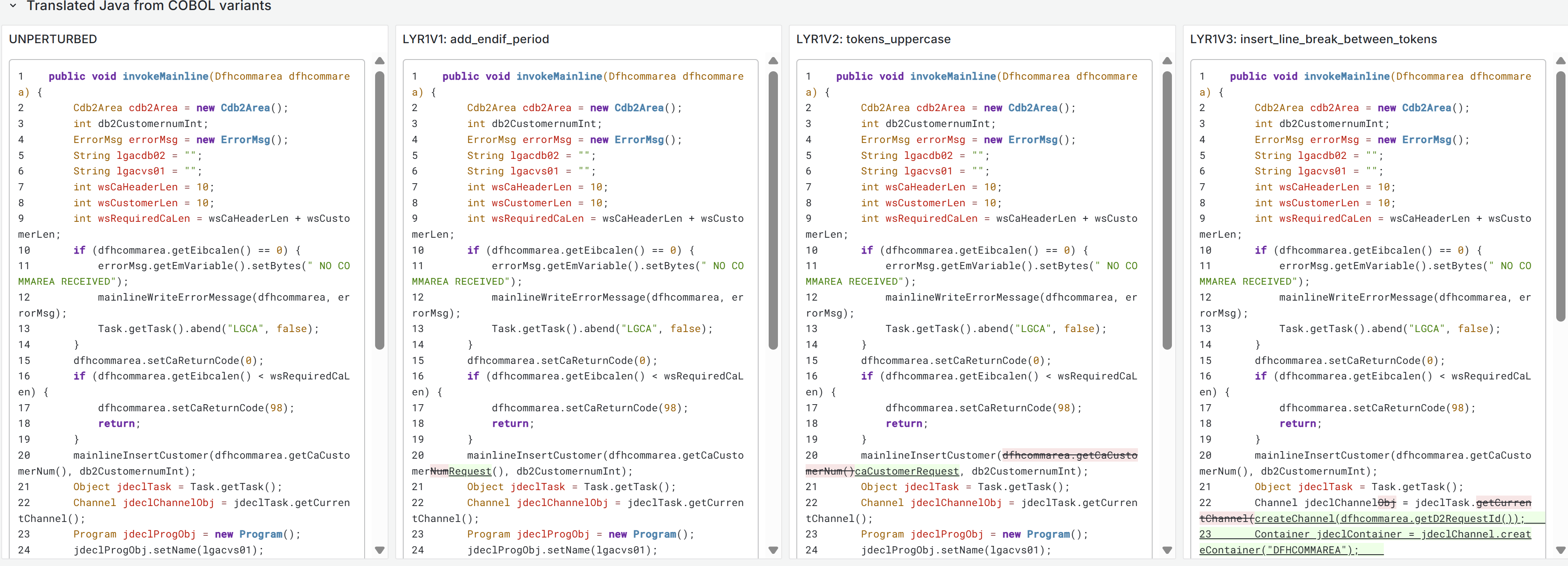}
  \caption{\label{fig:debug_dashboard}
  Partial screenshots of the debug dashboard.\\
  Top: Table of boolean metric values; red shading means there was a change relative to the original.\\
  Middle, bottom: Partial view of the input COBOL code and Java translations, with highlights showing the changes.
  }
\end{figure*}

